\UseRawInputEncoding 

\documentclass[]{article}

\usepackage{amsmath,amssymb}

\DeclareFontFamily{U}{musix}{}%
\DeclareFontShape{U}{musix}{m}{n}{%
  <-12>   musix11
  <12-15> musix13
  <15-18> musix16
  <18-23> musix20
  <23->   musix29
}{}%
\newcommand*\musix{\usefont{U}{musix}{m}{n}\selectfont}
\DeclareTextFontCommand{\textmusix}{\musix}

\usepackage[dvipdfmx]{graphicx}

\usepackage[top=30truemm,bottom=30truemm,left=25truemm,right=25truemm]{geometry}

\usepackage{amsthm}
\usepackage{bm}
\usepackage{color}
\usepackage{here}
\usepackage{comment}

\theoremstyle{plain}

\theoremstyle{definition}

\theoremstyle{remark}

\begin{document}

\title{General Theory of Music by Icosahedron 3: 
\\
\Large
Musical invariant and Melakarta raga}

\author{Yusuke Imai}
\date{%
\small
    Graduate School of Engineering Science, Osaka University, Toyonaka, Osaka 560-8531, Japan\\%
    \today
}

\maketitle

\small

\begin{center}
CONTACT: 93imaiyusuke@gmail.com
\end{center}

\begin{abstract}
Raga is a central musical concept in South Asia, especially India, and we investigate connections between Western classical music and Melakarta raga that is a raga in Karnatak (south Indian) classical music, through musical icosahedron. In our previous study, we introduced some kinds of musical icosahedra connecting various musical concepts in Western music: chromatic/whole tone musical icosahedra (type 1, type 2, type 3, type 4), Pythagorean/whole tone musical icosahedra (type 1', type 2', type 3', type 4'), and exceptional musical icosahedra (type ${\rm 1}^*$, type ${\rm 2}^*$, type ${\rm 3}^*$, type ${\rm 4}^*$). In this paper, first, we introduce kinds of musical icosahedra that connect the above musical icosahedra through two kinds of permutations of 12 tones: inter-permutations and intra-permutations, and we call them intermediate musical icosahedra. 

Next, we define a neighboring number as a number of pairs of neighboring two tones in a given scale that neighbor each other on a given musical icosahedron, and we also define a musical invariant as a linear combination of the neighboring numbers. We find there exists a pair of a musical invariant and scales that is constant for the type 1 (type 2) and the type 3 (type 4) and the intermediate musical icosahedra from the type 1 (type 2) to the type 3 (type 4), and for the type 1' (type 2') and the type 3' (type 4') and the intermediate musical icosahedra from the type 1' (type 2') to the type 3' (type 4'), and for the type ${\rm 1}^*$ (type ${\rm 2}^*$) and the type ${\rm 3}^*$ (type ${\rm 4}^*$) and the intermediate musical icosahedra from the type ${\rm 1}^*$ (type ${\rm 2}^*$) to the type ${\rm 3}^*$ (type ${\rm 4}^*$), and for the intermediate musical icosahedra from the type n to the type n'. We analyze their mathematical structure and show the results by using tables and diagrams.

Last, we define an extension of a given scale by the inter-permutations of a given musical icosahedron: the permutation-extension. Then, we show that the permutation-extension of the $C$ major scale by Melakarta raga musical icosahedra that are four of the intermediate musical icosahedra from the type 1 chromatic/whole tone musical icosahedron to the type 1' Pythagorean/whole tone musical icosahedron, is a set of all the scales included in Melakarta raga. There exists a musical invariant that is constant for all the musical icosahedra corresponding to the scales of Melakarta raga, and we obtained a diagram representation of those scales characterizing the musical invariant.
\end{abstract}

\newpage
\section{Introduction}
Music may exist everywhere and everywhen, and then it is interesting to distinguish them and investigate connections among them. In the previous papers (Imai, Dellby, Tanaka, 2021 and Imai, 2021), we obtained many connections among musical concepts in Western music through mathematical properties of the regular icosahedron (graph geometry, golden ratio, symmetry): ``artificial" scales (chromatic scale and whole tone scales) and ``natural" scales (major/minor scales, Gregorian modes, hexatonic major/minor scales), chromatic scale and Pythagorean chain (succession of the fifth), the famous prelude composed by Johann Sebastian Bach (BWV 846) and figures characterized by the golden ratio (a golden triangle/gnomon/rectangle).

In this paper, we focus on raga that is a central element of South Asian classical music (Powers and Widdess, 2020) and investigate a relation between Western music and South Asian music. Powers and Widdess summarized raga as follows. The word, raga, originates from ``$ra\tilde{n}j$, `to be coloured, to redden', hence `to be affected, moved, charmed, delighted'". Raga is not a mere musical concept, but ``Aesthetic and extra-musical aspects of raga have been, and remain, profoundly important in Indian culture. They have included supernatural powers over the physical universe and associations with particular divinities, human characters, seasons, and times of day." Also, a raga is described by (i) ``its melodic features, that is, in terms of successions of scale degrees" and (ii) ``by its melodic features, that is, in terms of successions of scale degrees".

There are two important kinds of raga: raga in Hindustani (north Indian) classical music and raga in Karnatak (south Indian) classical music. Raga has been studied enthusiastically from various viewpoints. For example, Krishna attempted to ``define the abstract concept of raga in its entirety within the aesthetics" (Krishna, 2012). Also, a relation between emotion and Hindustani raga music was analyzed by Mathur \emph{et al} (Mathur \emph{et al}., 2015). A relationship between raga and sleep architecture was investigated by Gitanjali(Gitanjali, 1998). Sairam studied Indian music as a music therapy (Sairam, 2006). Pandey, Mishra, and Ipe studied raga from the viewpoint of computational musicology (Pandey, Mishra, Ipe, 2003). Sridhar and Geetha presented a way to identify raga from a Carnatic music (Sridhar, Geetha, 2009) and data mining of raga for Indian music was studied by Shetty and Achary (Shetty, Achary, 2009). Ross and Rao proposed a way to detect raga from hindustani classical music (Ross and Rao, 2012). Rao \emph{et al}. applied ``Dynamic time warping and HMM based classification on time series of detected pitch values used for the melodic representation of a phrase" (Rao \emph{et al}., 2014).

This paper especially deals with the Melakarta raga that is a raga in Karnatak classical music and is characterized by fundamental 72 modes. By using $C$, $C\sharp$, $D$, $E\flat$, $E$, $F$, $F\sharp$, $G$, $A$, $B\flat$, $B$, they are represented as follows. 
\\
\\
\indent
1: Kanakangi: $C$, $C\sharp$, $D$, $F$, $G$, $G\sharp$, $A$,

2: Ratnangi: $C$, $C\sharp$, $D$, $F$, $G$, $G\sharp$, $B\flat$,

3: Ganamurti: $C$, $C\sharp$, $D$, $F$, $G$, $G\sharp$, $B$,

4: Vanaspati: $C$, $C\sharp$, $D$, $F$, $G$, $A$, $B\flat$,

5: Manavati: $C$, $C\sharp$, $D$, $F$, $G$, $A$, $B$,

6: Tanarupi: $C$, $C\sharp$, $D$, $F$, $G$, $B\flat$, $B$,

7: Senavati: $C$, $C\sharp$, $E\flat$, $F$, $G$, $G\sharp$, $A$,

8: Hanumatodi: $C$, $C\sharp$, $E\flat$, $F$, $G$, $G\sharp$, $B\flat$,

9: Dhenuka: $C$, $C\sharp$, $E\flat$, $F$, $G$, $G\sharp$, $B$,

10: Natakapriya: $C$, $C\sharp$, $E\flat$, $F$, $G$, $A$, $B\flat$,

11: Kokilapriya: $C$, $C\sharp$, $E\flat$, $F$, $G$, $A$, $B$,

12: Rupavati: $C$, $C\sharp$, $E\flat$, $F$, $G$, $B\flat$, $B$,

13: Gayakapriya: $C$, $C\sharp$, $E$, $F$, $G$, $G\sharp$, $A$,

14: Vakulabharanam: $C$, $C\sharp$, $E$, $F$, $G$, $G\sharp$, $B\flat$,

15: Mayamalavagowla: $C$, $C\sharp$, $E$, $F$, $G$, $G\sharp$, $B$,

16: Chakravakam: $C$, $C\sharp$, $E$, $F$, $G$, $A$, $B\flat$,

17: Suryakantam: $C$, $C\sharp$, $E$, $F$, $G$, $A$, $B$,

18: Hatakambari: $C$, $C\sharp$, $E$, $F$, $G$, $B\flat$, $B$,

19: Jhankaradhwani: $C$, $D$, $E\flat$, $F$, $G$, $G\sharp$, $A$,

20: Natabhairavi: $C$, $D$, $E\flat$, $F$, $G$, $G\sharp$, $B\flat$,

21: Keeravani: $C$, $D$, $E\flat$, $F$, $G$, $G\sharp$, $B$,

22: Kharaharapriya: $C$, $D$, $E\flat$, $F$, $G$, $A$, $B\flat$,

23: Gourimanohari: $C$, $D$, $E\flat$, $F$, $G$, $A$, $B$,

24: Varunapriya: $C$, $D$, $E\flat$, $F$, $G$, $B\flat$, $B$,

25: Mararanjani: $C$, $D$, $E$, $F$, $G$, $G\sharp$, $A$,

26: Charukesi: $C$, $D$, $E$, $F$, $G$, $G\sharp$, $B\flat$,

27: Sarasangi: $C$, $D$, $E$, $F$, $G$, $G\sharp$, $B$,

28: Harikambhoji: $C$, $D$, $E$, $F$, $G$, $A$, $B\flat$,

29: Dheerasankarabaranam: $C$, $D$, $E$, $F$, $G$, $A$, $B$,

30: Naganandini: $C$, $D$, $E$, $F$, $G$, $B\flat$, $B$,

31: Yagapriya: $C$, $E\flat$, $E$, $F$, $G$, $G\sharp$, $A$,

32: Ragavardhini: $C$, $E\flat$, $E$, $F$, $G$, $G\sharp$, $B\flat$,

33: Gangeyabhushani: $C$, $E\flat$, $E$, $F$, $G$, $G\sharp$, $B$,

34: Vagadheeswari: $C$, $E\flat$, $E$, $F$, $G$, $A$, $B\flat$,

35: Shulini: $C$, $E\flat$, $E$, $F$, $G$, $A$, $B$,

36: Chalanata: $C$, $E\flat$, $E$, $F$, $G$, $B\flat$, $B$,

37: Salagam: $C$, $C\sharp$, $D$, $F\sharp$, $G$, $G\sharp$, $A$,

38: Jalarnavam: $C$, $C\sharp$, $D$, $F\sharp$, $G$, $G\sharp$, $B\flat$,

39: Jhalavarali: $C$, $C\sharp$, $D$, $F\sharp$, $G$, $G\sharp$, $B$,

40: Navaneetam: $C$, $C\sharp$, $D$, $F\sharp$, $G$, $A$, $B\flat$,

41: Pavani: $C$, $C\sharp$, $D$, $F\sharp$, $G$, $A$, $B$,

42: Raghupriya: $C$, $C\sharp$, $D$, $F\sharp$, $G$, $B\flat$, $B$,

43: Gavambhodi: $C$, $C\sharp$, $E\flat$, $F\sharp$, $G$, $G\sharp$, $A$,

44: Bhavapriya: $C$, $C\sharp$, $E\flat$, $F\sharp$, $G$, $G\sharp$, $B\flat$,

45: Shubhapantuvarali: $C$, $C\sharp$, $E\flat$, $F\sharp$, $G$, $G\sharp$, $B$,

46: Shadvidamargini: $C$, $C\sharp$, $E\flat$, $F\sharp$, $G$, $A$, $B\flat$,

47: Suvarnangi: $C$, $C\sharp$, $E\flat$, $F\sharp$, $G$, $A$, $B$,

48: Divyamani: $C$, $C\sharp$, $E\flat$, $F\sharp$, $G$, $B\flat$, $B$,

49: Dhavalambari: $C$, $C\sharp$, $E$, $F\sharp$, $G$, $G\sharp$, $A$,

50: Namanarayani: $C$, $C\sharp$, $E$, $F\sharp$, $G$, $G\sharp$, $B\flat$,

51: Kamavardhini: $C$, $C\sharp$, $E$, $F\sharp$, $G$, $G\sharp$, $B$,

52: Ramapriya: $C$, $C\sharp$, $E$, $F\sharp$, $G$, $A$, $B\flat$,

53: Gamanashrama: $C$, $C\sharp$, $E$, $F\sharp$, $G$, $A$, $B$,

54: Vishwambari: $C$, $C\sharp$, $E$, $F\sharp$, $G$, $B\flat$, $B$,

55: Shamalangi: $C$, $D$, $E\flat$, $F\sharp$, $G$, $G\sharp$, $A$,

56: Shanmukhapriya: $C$, $D$, $E\flat$, $F\sharp$, $G$, $G\sharp$, $B\flat$,

57: Simhendramadhyamam: $C$, $D$, $E\flat$, $F\sharp$, $G$, $G\sharp$, $B$,

58: Hemavati: $C$, $D$, $E\flat$, $F\sharp$, $G$, $A$, $B\flat$,

59: Dharmavati: $C$, $D$, $E\flat$, $F\sharp$, $G$, $A$, $B$,

60: Neetimati: $C$, $D$, $E\flat$, $F\sharp$, $G$, $B\flat$, $B$,

61: Kantamani: $C$, $D$, $E$, $F\sharp$, $G$, $G\sharp$, $A$,

62: Rishabhapriya: $C$, $D$, $E$, $F\sharp$, $G$, $G\sharp$, $B\flat$,

63: Latangi: $C$, $D$, $E$, $F\sharp$, $G$, $G\sharp$, $B$,

64: Vachaspati: $C$, $D$, $E$, $F\sharp$, $G$, $A$, $B\flat$,

65: Mechakalyani: $C$, $D$, $E$, $F\sharp$, $G$, $A$, $B$,

66: Chitrambari: $C$, $D$, $E$, $F\sharp$, $G$, $B\flat$, $B$,

67: Sucharitra: $C$, $E\flat$, $E$, $F\sharp$, $G$, $G\sharp$, $A$,

68: Jyoti swarupini: $C$, $E\flat$, $E$, $F\sharp$, $G$, $G\sharp$, $B\flat$,

69: Dhatuvardani: $C$, $E\flat$, $E$, $F\sharp$, $G$, $G\sharp$, $B$,

70: Nasikabhushani: $C$, $E\flat$, $E$, $F\sharp$, $G$, $A$, $B\flat$,

71: Kosalam: $C$, $E\flat$, $E$, $F\sharp$, $G$, $A$, $B$,

72: Rasikapriya: $C$, $E\flat$, $E$, $F\sharp$, $G$, $B\flat$, $B$.
\\
\\
\indent
Note that all the scales include $C$ and $G$, and the scale 1-36 (37-72) include $F$ ($F\sharp$) and do not include $F\sharp$ ($F$), and the scale 1-6 and 37-42 include $C\sharp$ and $D$, and do not include $E\flat$ and $E$, and the scale 7-12 and 43-48 include $C\sharp$ and $E\flat$, and do not include $D$ and $E$, and the scale 13-18 and 49-54 include $C\sharp$ and $E$, and do not include $D$ and $E\flat$, and the scale 19-24 and 55-60 include $D$ and $E\flat$, and do not include $C\sharp$ and $E$, and the scale 25-30 and 61-66 include $D$ and $E$, and do not include $C\sharp$ and $E\flat$, and the scale 31-36 and 67-72 include $E\flat$ and $E$, and do not include $C\sharp$ and $D$, and the scale 1, 7, 13, 19, 25, 31, 37, 43, 49, 55, 61, 67 include $G\sharp$ and $A$ and do not include $B\flat$ and $B$, and the scale 2, 8, 14, 20, 26, 32, 38, 44, 50, 56, 62, 68 include $G\sharp$ and $B\flat$ and do not include $A$ and $B$, and the scale 3, 9, 15, 21, 27, 33, 39, 45, 51, 57, 63, 69 include $G\sharp$ and $B$ and do not include $A$ and $B\flat$, and the scale 4, 10, 16, 22, 28, 34, 40, 46, 52, 58, 64, 70 include $A$ and $B\flat$ and do not include $G\sharp$ and $B$, and the scale 5, 11, 17, 23, 29, 35, 41, 47, 53, 59, 65, 71 include $A$ and $B$ and do not include $G\sharp$ and $B\flat$, and the scale 6, 12, 18, 24, 30, 36, 42, 48, 54, 60, 66, 72 include $B\flat$ and $B$ and do not include $G\sharp$ and $A$. Then, the Melakarta raga is obtained by choosing $C$ and $G$, and one tone from $F$ and $F\sharp$, two tones from $C\sharp$, $D$, $E\flat$, $E$, and two tones from $G\sharp$, $A$, $B\flat$, $B$. Our goal is to derive this structure by musical icosahedron.

We investigate a connection between musical concepts in Western music and the structure of the Melakarta raga through musical icosahedron. Concretely, we define the following new concepts: intra-permutation, inter-permutation,  intermediate musical icosahedron, neighboring number, and musical invariant. This theory can be regarded as a discrete analogue of fiber bundle theory and then, we obtain these definitions and concepts in Western music naturally lead to the structure of Melakarta raga. This paper is organized as follows. In Sec.\,II, we define some kinds of permutations of 12 tones, intermediate musical icosahedron, musical invariant, and analyze their structure. In Sec.\,III, we propose a way of extending a given scale by using the concepts defined in Sec,\,II and investigate a connection between Western music and Melakarta raga.

\newpage
\section{Parmuatation, Intermediate Musical Icosahedron, and Musical Invariant}
In this section, we introduce the following concepts, inter-permutation, intra-permutation, intermediate musical icosahedron and musical invariant. A theory presented here can be regarded as a kind of discrete analogues of the fiber bundle theory.

\subsection{Inter-permutation}
\begin{figure}[H]
\centering
{%
\resizebox*{10cm}{!}{\includegraphics{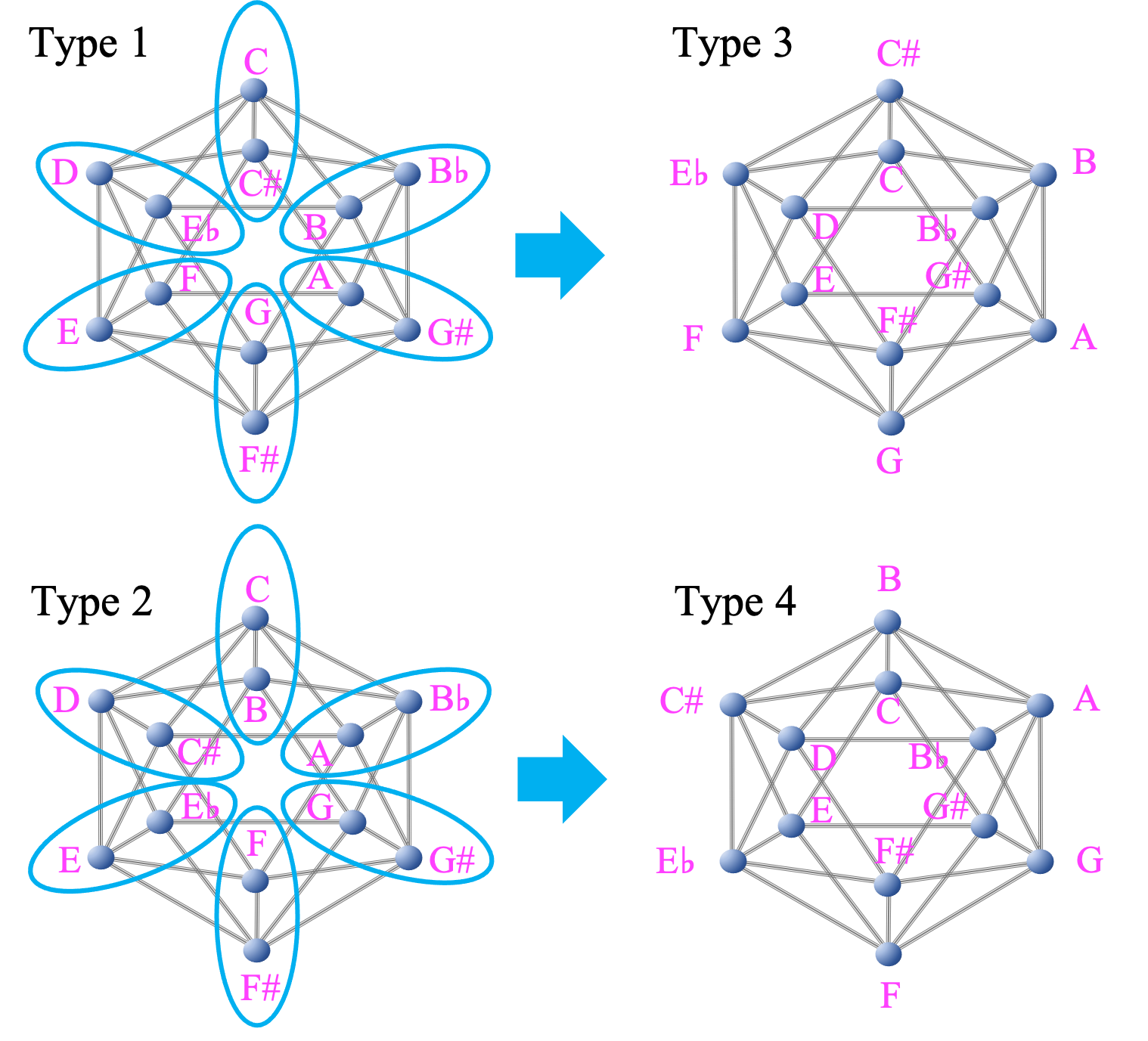}}}\hspace{5pt}
\caption{The type 3 (type 4) is obtained by applying the permutations represented by the sky-blue ellipses to the type 1 (type 2).} \label{perm1}
\end{figure}

\begin{figure}[H]
\centering
{%
\resizebox*{10cm}{!}{\includegraphics{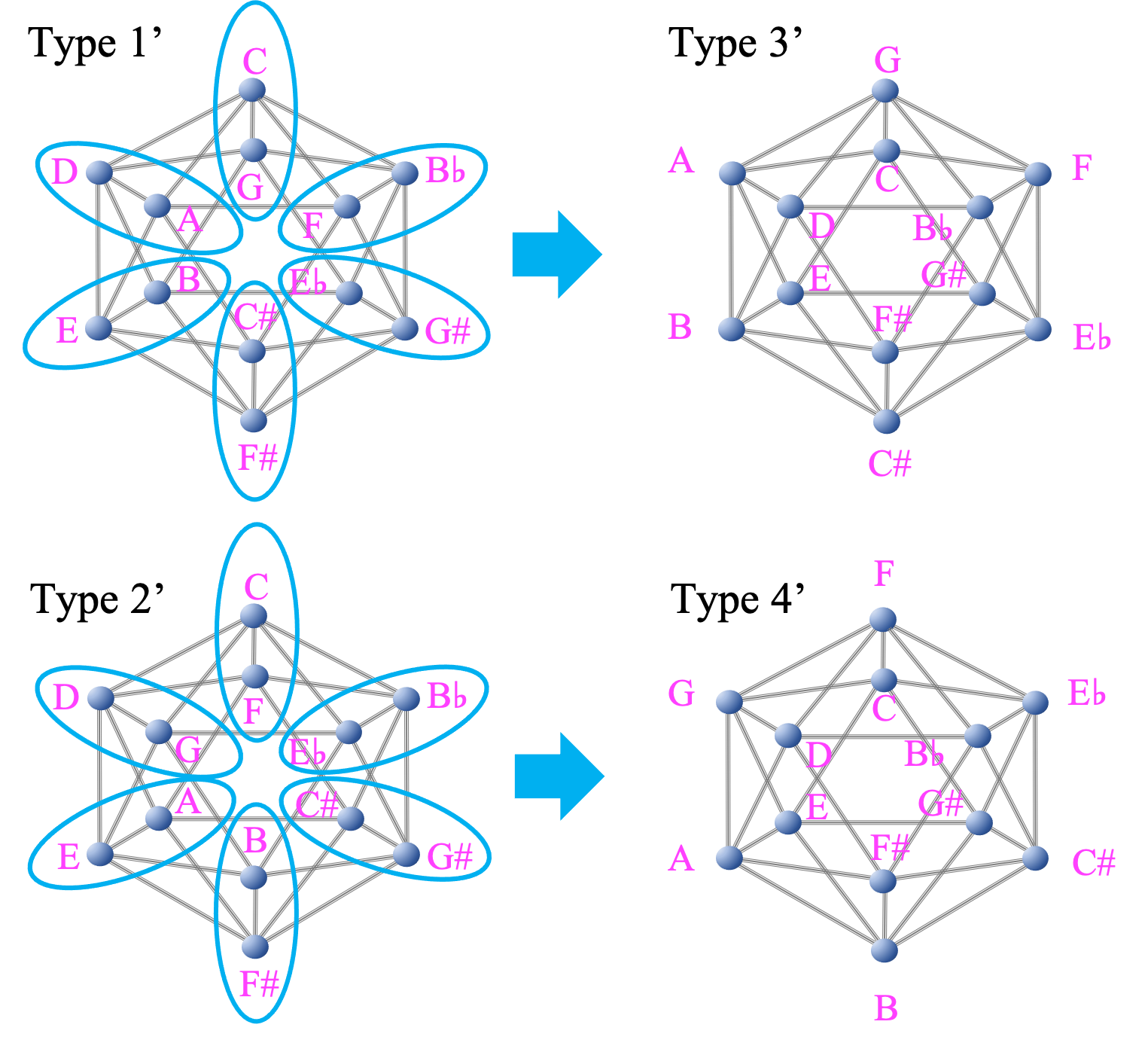}}}\hspace{5pt}
\caption{The type 3' (type 4') is obtained by applying the permutations represented by the sky-blue ellipses to the type 1' (type 2').} \label{perm2}
\end{figure}

\begin{figure}[H]
\centering
{%
\resizebox*{10cm}{!}{\includegraphics{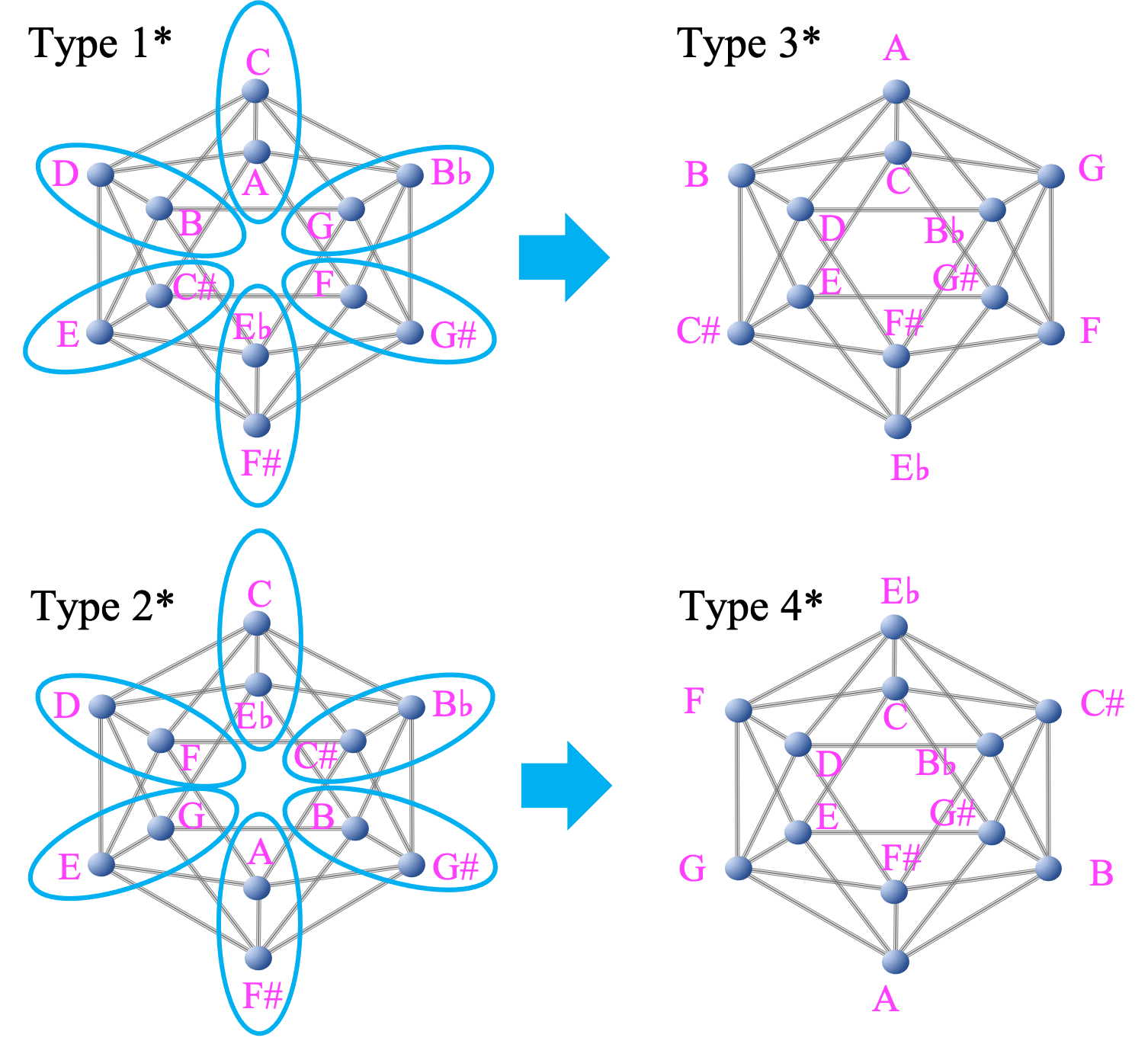}}}\hspace{5pt}
\caption{The type ${\rm 3^*}$ (type ${\rm 4^*}$ ) is obtained by applying the permutations represented by the sky-blue ellipses to the type ${\rm 1^*}$  (type ${\rm 2^*}$ ).} \label{perm3}
\end{figure}

\begin{figure}[H]
\centering
{%
\resizebox*{5cm}{!}{\includegraphics{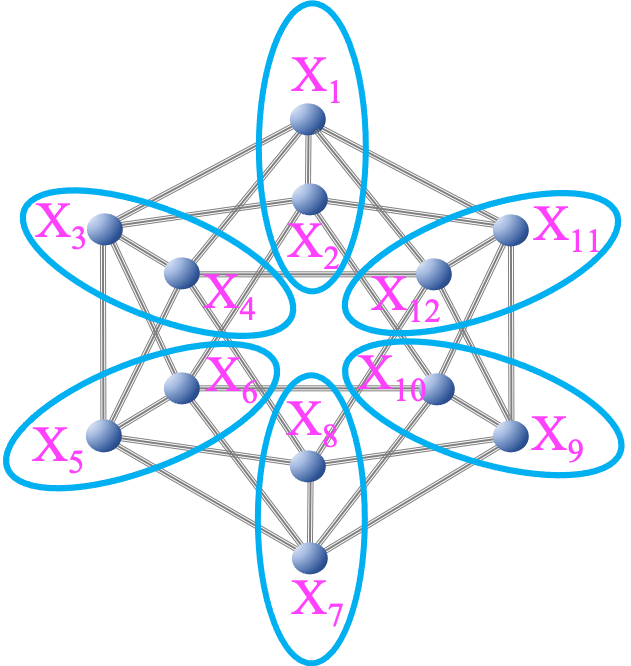}}}\hspace{5pt}
\caption{The inter-permutations for a musical icosahedron with $X_1$, $X_2$, $X_3$, $X_4$, $X_5$, $X_6$, $X_7$, $X_8$, $X_9$, $X_{10}$, $X_{11}$, $X_{12}$.} \label{perm4}
\end{figure}

In our first paper (Imai, Dellby, Tanaka, 2021), we proposed some kinds of musical icosahedra: chromatic/whole tone musical icosahedra, Pythagorean/whole tone musical icosahedra, exceptional musical icosahedra. All the kinds of musical icosahedra share many characteristics, and one of them is a relation among the types of each kind of musical icosahedra. For example, the type 3 is obtained by the following permutations for the type 1: $C\leftrightarrow C\sharp$, $D\leftrightarrow E\flat$, $E\leftrightarrow F$, $F\sharp \leftrightarrow G$, $G\sharp \leftrightarrow A$, $B\flat \leftrightarrow B$. The type 4 is obtained by the following permutations for the type 2: $B\leftrightarrow C$, $C\sharp\leftrightarrow D$, $E\flat \leftrightarrow E$, $F \leftrightarrow F\sharp$, $G \leftrightarrow G\sharp$, $A \leftrightarrow B\flat$. The type 3' is obtained by the following permutations for the type 1': $C\leftrightarrow G$, $D\leftrightarrow A$, $E\leftrightarrow B$, $F\sharp \leftrightarrow C\sharp$, $G\sharp \leftrightarrow E\flat$, $B\flat \leftrightarrow F$. The type 4' is obtained by the following permutations for the type 2': $F\leftrightarrow C$, $G\leftrightarrow D$, $A \leftrightarrow E$, $B \leftrightarrow F\sharp$, $C\sharp \leftrightarrow G\sharp$, $E\flat \leftrightarrow B\flat$. The type ${\rm 3^*}$ is obtained by the following permutations for the type ${\rm 1^*}$: $C\leftrightarrow A$, $D\leftrightarrow B$, $E\leftrightarrow C\sharp$, $F\sharp \leftrightarrow E\flat$, $G\sharp \leftrightarrow F$, $B\flat \leftrightarrow G$. The type ${\rm 4^*}$ is obtained by the following permutations for the type ${\rm 2^*}$: $E\flat\leftrightarrow C$, $F\leftrightarrow D$, $G \leftrightarrow E$, $A \leftrightarrow F\sharp$, $B \leftrightarrow G\sharp$, $C\sharp \leftrightarrow B\flat$. These results are summarized in Fig.\,\ref{perm1}, Fig.\,\ref{perm2}, and Fig.\,\ref{perm3}. We define an inter-permutation for a given musical icosahedron as a permutation of two tones corresponding to two vertices represented by a sky-blue ellipse in Fig.\,\ref{perm4}.

\begin{figure}[H]
\centering
{%
\resizebox*{10cm}{!}{\includegraphics{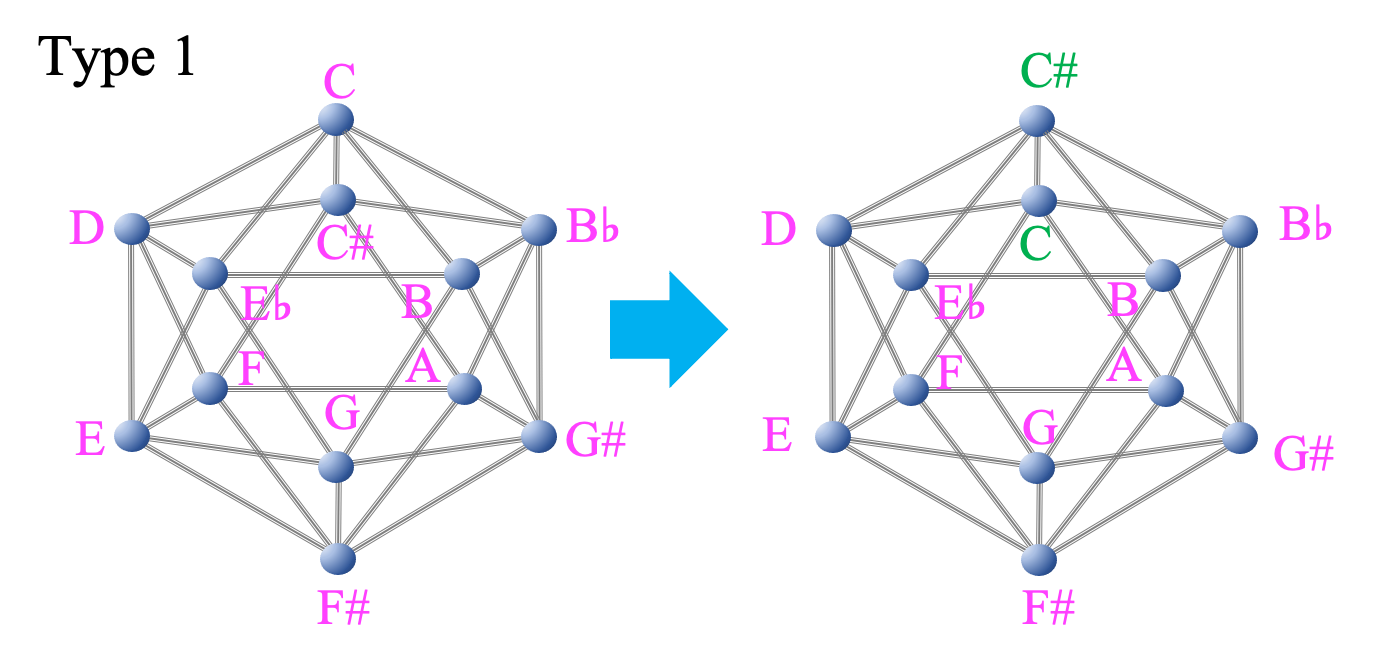}}}\hspace{5pt}
\caption{A musical icosahedron obtained by applying one of the inter-permutations shown in Fig.\,\ref{perm1} to the type 1.} \label{perm_one}
\end{figure}

Now, we consider a musical icosahedron that is obtained by applying one of the inter-permutations shown in Fig.\,\ref{perm1} to the type 1. For example, we obtain a musical icosahedron shown in Fig.\,\ref{perm_one}. This musical icosahedron satisfies the neighboring condition for the whole tone scale including $C$ while does not satisfy the neighboring for the chromatic scale because $B$ is not connected to $C$. Then, this musical icosahedron seems not to be related to the type 1 closely.

However, by introducing some mathematical definitions, one can understand that these two musical icosahedra share the important characteristic. First, we introduce a concept, neighboring number. The neighboring number $N(A, X)$, for a given scale $A$ and a musical icosahedron $X$, is defined as the number of pairs of neighboring two tones in $A$ whose corresponding vertices are connected in $X$. Obviously, for a scale $A$ constructed by $n$ tones, a musical icosahedron $X$, $N(A, X)=n$ if and only if $X$ satisfies the neighboring condition for $A$. Therefore, the neighboring number is a concept that extends the neighboring condition. Also, for any natural number $k$, $n_1,\cdots, n_k$ and any scales, $A_1,  \cdots,  A_k$, we define musical invariant as $N(A_1, \cdots, A_k, n_1, \cdots, n_k, X) = \sum_{j = 1}^k n_j N(A_j, X)$.

\begin{figure}[H]
\centering
{%
\resizebox*{14cm}{!}{\includegraphics{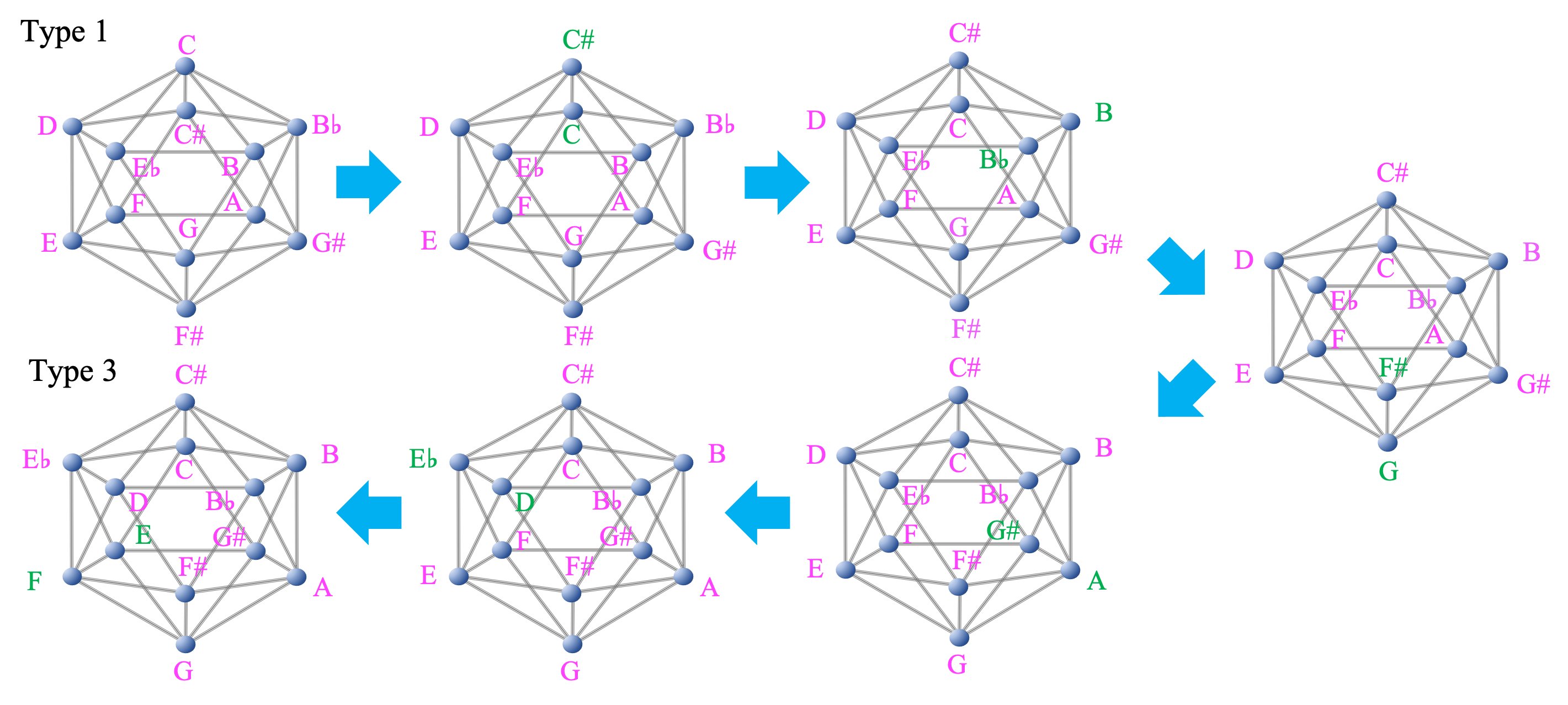}}}\hspace{5pt}
\caption{A succession of the intermediate musical icosahedra from the type 1 to the type 3. For all the musical icosahedra, $N(C, W, 1, 1/2, \cdot \ )=15$.} \label{perm_suc}
\end{figure}

Remember that the neighboring number for the whole tone scale including $C\sharp$ is $0$ in the type 1. Hence, $N(C, {\rm type \ 1})=12$ and $N(W, {\rm type \ 1})=6$ where $C$ means the chromatic scale and $W$ means the whole tone scales\footnote{$N(W, {\rm type \ 1})\equiv N(W1, {\rm type \ 1}) + N(W2, {\rm type \ 1})$ where $W1 = C, D, E, F\sharp, G\sharp, B\flat$ and $W2 = C\sharp, E\flat, F, G, A, B$.}. Also, the neighboring number for the whole tone scale including $C\sharp$ is $2$ in the right musical icosahedron shown in Fig.\,\ref{perm_one}. Then, $N(C, W, 1, 1/2, \ \cdot \ ) = 15$ for the two musical icosahedra shown in Fig.\,\ref{perm_one}. If one applies any combination of the permutations shown in Fig.\,\ref{perm1} to the type 1, the resulting icosahedron $X$ satisfies $N(C, W, 1, 1/2, X) = 15$. We call these 62 ($=2^6-2$) musical icosahedra the intermediate musical icosahedra from the type 1 to the type 3. A succession of the intermediate musical icosahedra from the type 1 to the type 3 shown in Fig.\,\ref{perm_suc} connects the type 1 with the type 3 preserving $N(C, W, 1, 1/2, \ \cdot \ )$. This is analogous to transformations of a compact two-dimensional Riemannian manifold preserving the number of holes such as a continuous transformation from a coffee cup to a donut.

Intermediate musical icosahedra can be defined also from the type 3 to the type 1, from the type 2 (type 4) to the type 4 (type 2), from the type 1' (type 3') to the type 3' (type 1'), the type 2' (type 4') and the type 4' (type 2'), from the type ${\rm 1}^*$ (type ${\rm 3}^*$) to the type ${\rm 3}^*$ (type ${\rm 1}^*$), from the type ${\rm 2}^*$ (type ${\rm 4}^*$) to the type ${\rm 4}^*$ (type ${\rm 2}^*$). Now, we define the following six scales:
\\
\\
\indent
(+1, -11) scale: $C$, $C\sharp$, $D$, $E\flat$, $E$, $F$, $F\sharp$, $G$, $G\sharp$, $A$, $B\flat$, $B$,

(+3, -1) scale: $C$, $E\flat$, $D$, $F$, $E$, $G$, $F\sharp$, $A$, $G\sharp$, $B$, $B\flat$, $C\sharp$,

(+5, -3) scale: $C$, $F$, $D$, $G$, $E$, $A$, $F\sharp$, $B$, $G\sharp$, $C\sharp$, $B\flat$, $E\flat$,

(+7, -5) scale: $C$, $G$, $D$, $A$, $E$, $B$, $F\sharp$, $C\sharp$, $G\sharp$, $E\flat$, $B\flat$, $F$,

(+9, -7) scale: $C$, $A$, $D$, $B$, $E$, $C\sharp$, $F\sharp$, $E\flat$, $G\sharp$, $F$, $B\flat$, $G$,

(+11, -9) scale: $C$, $B$, $D$, $C\sharp$, $E$, $E\flat$, $F\sharp$, $F$, $G\sharp$, $G$, $B\flat$, $A$.
\\
\\
\indent
The (+1, -11) scale (chromatic scale) in the previous discussion should be replaced by the (+3, -1) scale when dealing with the intermediate musical icosahedra from the type 3 to the type 1, by the (+11, -9) scale when dealing with the intermediate musical icosahedra from the type 2 to the type 4, by the (+7, -5) scale when dealing with the intermediate musical icosahedra from the type 1' to the type 3', by the (+9, -7) scale when dealing with the intermediate musical icosahedra from the type 3' to the type 1', by the (+5, -3) scale when dealing with the intermediate musical icosahedra from the type 2' to the type 4', by the (+7, -5) scale when dealing with the intermediate musical icosahedra from the type 4' to the type 2', by the (+9, -7) scale when dealing with the intermediate musical icosahedra from the type ${\rm 1^*}$ to the type ${\rm 3^*}$, by the (+11, -1) scale when dealing with the intermediate musical icosahedra from the type ${\rm 3^*}$ to the type ${\rm 1^*}$, by the (+3, -1) scale when dealing with the intermediate musical icosahedra from the type ${\rm 2^*}$ to the type ${\rm 4^*}$, by the (+5, -3) scale when dealing with the intermediate musical icosahedra from the type ${\rm 4^*}$ to the type ${\rm 2^*}$.

\begin{figure}[H]
\centering
{%
\resizebox*{14cm}{!}{\includegraphics{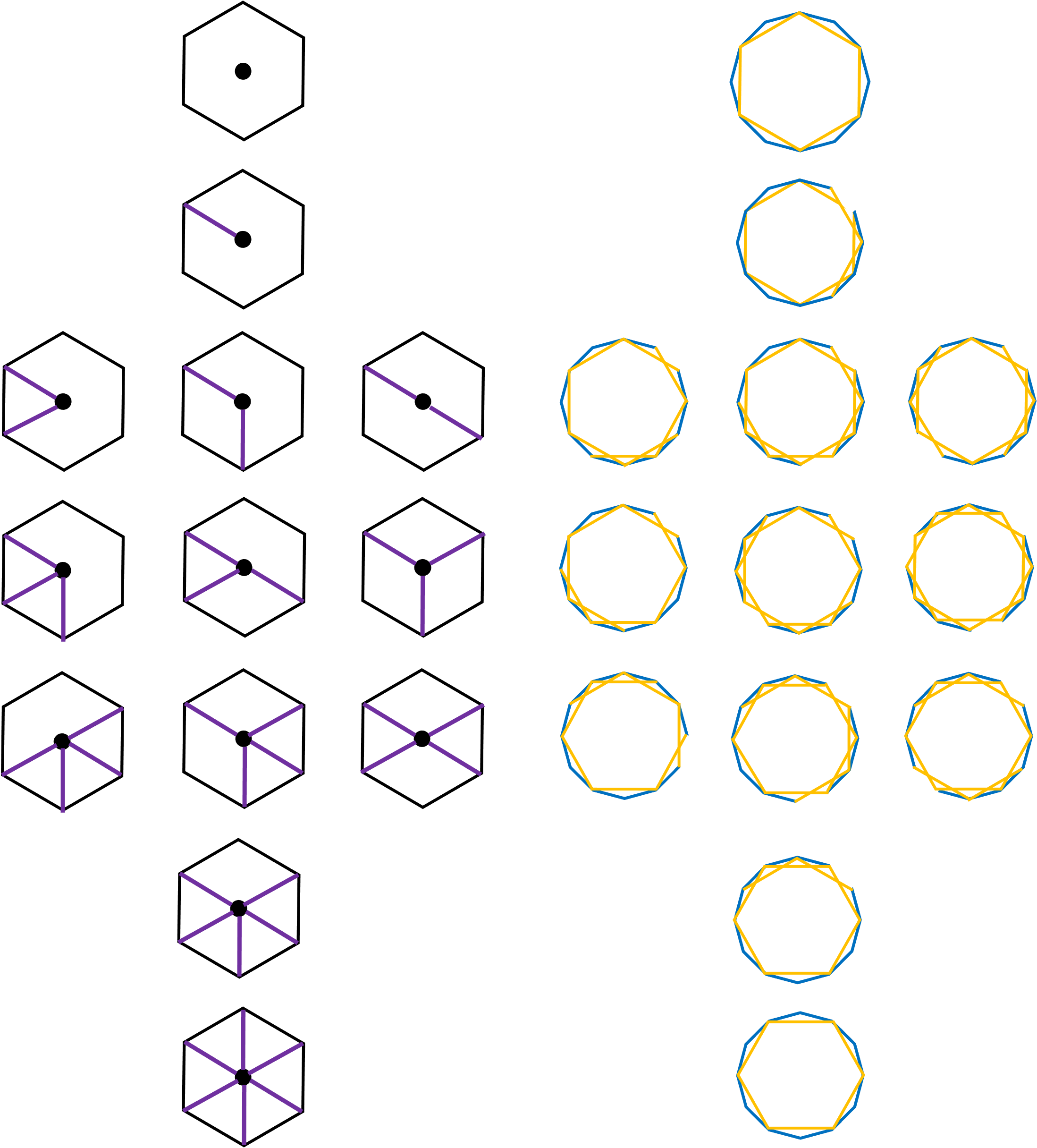}}}\hspace{5pt}
\caption{A schematic representation of the structure of the intermediate musical icosahedra. The diagrams on the left side show a kind of the inter-permutations. The diagrams on the right side show the neighboring property of the (+1, -11) scale, (+3, -1) scale, (+5, -3) scale, (+7, -5) scale, (+9, -7) scale, and (+11, -9) scale.} \label{structure_diagram}
\end{figure}

\begin{figure}[H]
\centering
{%
\resizebox*{14cm}{!}{\includegraphics{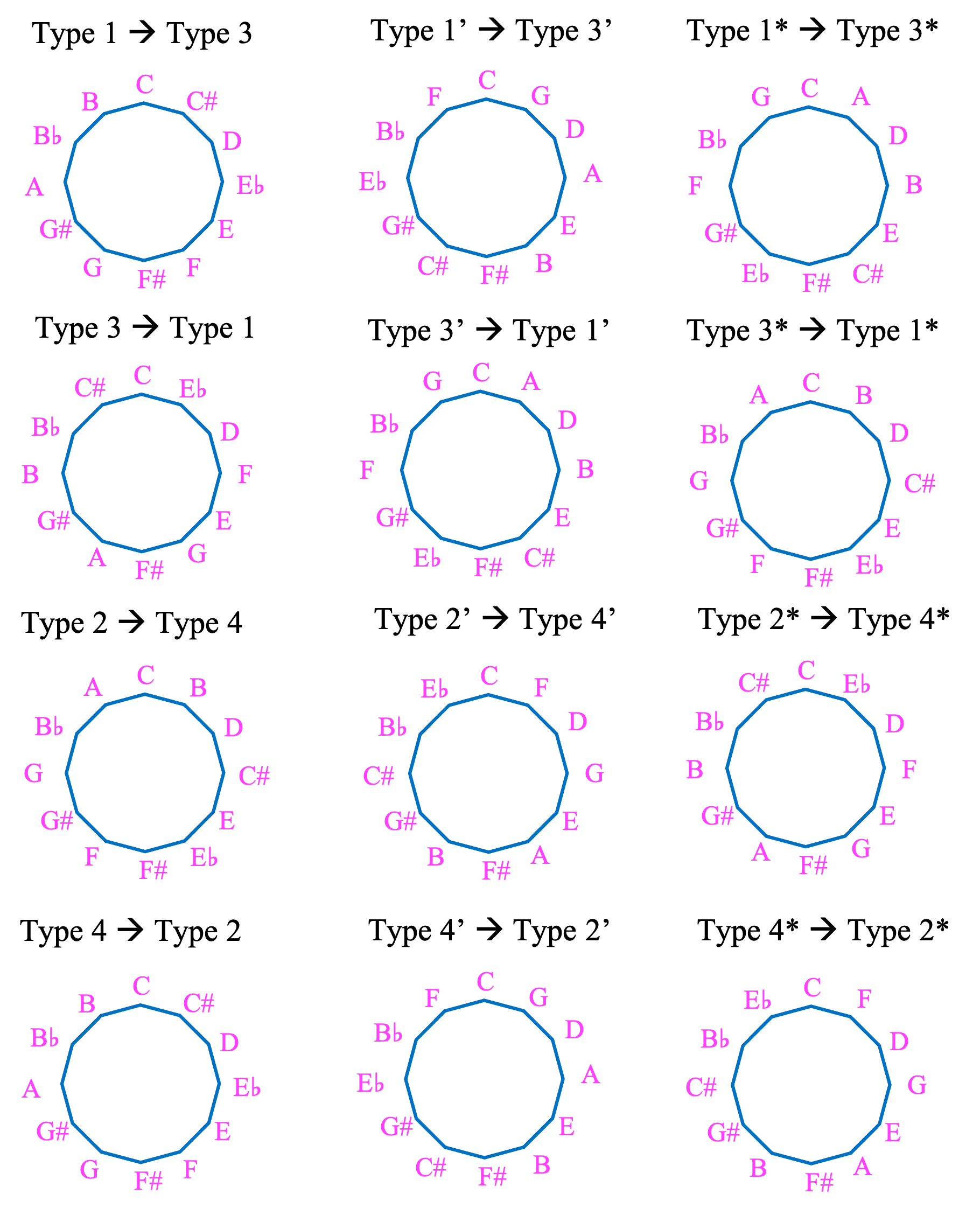}}}\hspace{5pt}
\caption{Arrangements of 12 tones on the regular dodecagon for all the kinds of the intermediate musical icosahedra.} \label{12}
\end{figure}

All the intermediate musical icosahedra have the same structure. By considering symmetry, we have 13 kinds of musical icosahedra representing the structure of the intermediate musical icosahedra, and they are schematically shown in Fig.\,\ref{structure_diagram}. We explain what these diagrams represent. The figures on the right side show the neighboring property of the (+1, -11) scale, (+3, -1) scale, (+5, -3) scale, (+7, -5) scale, (+9, -7) scale, and (+11, -9) scale. Depending on a choice of the initial musical icosahedron, an arrangement of 12 tones on the vertices of the regular dodecagon changes (Fig.\,\ref{12}).

\begin{figure}[H]
\centering
{%
\resizebox*{10cm}{!}{\includegraphics{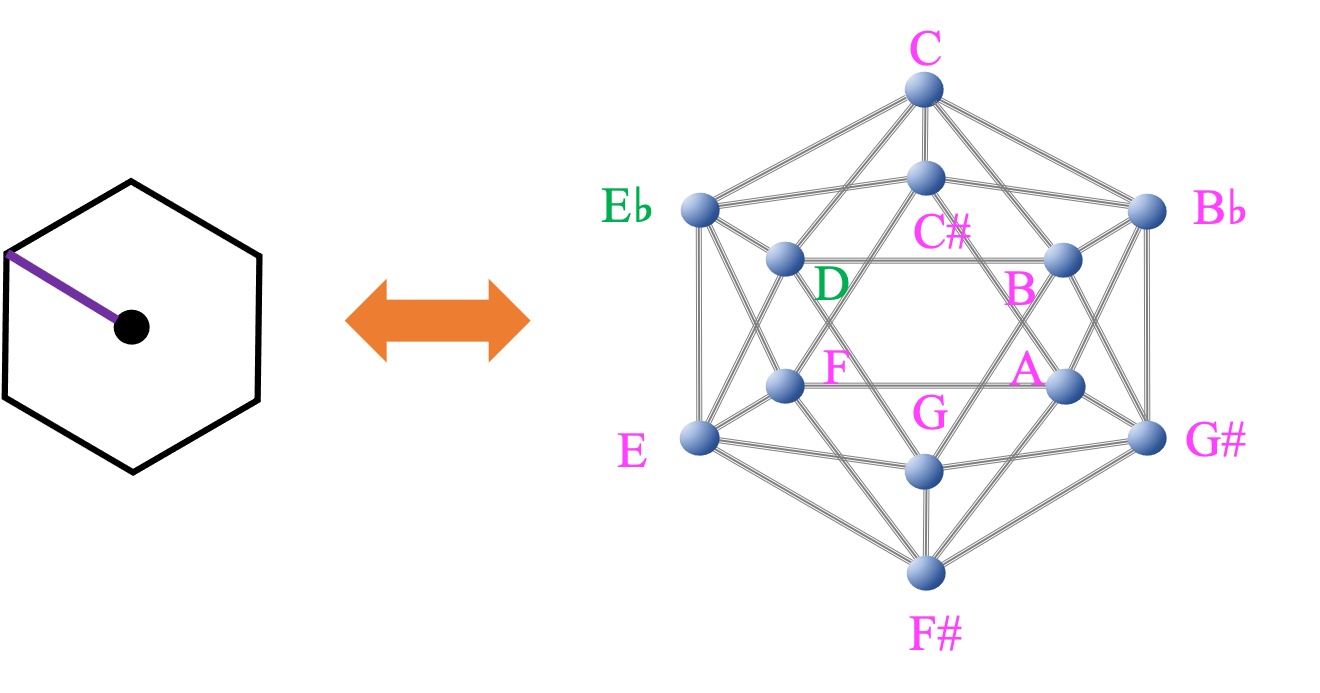}}}\hspace{5pt}
\caption{The schematic representation on the left side corresponds to the intermediate musical icosahedron on the right side in case of dealing with the intermediate icosahedra from the type 1 to the type 3.} \label{perm_schem}
\end{figure}

Also, the schematic representations on the left side show a kind of the inter-permutations. For example, the diagram on the second row in Fig.\,\ref{structure_diagram} represents an inter-permutation from the type 1 to the type 3 shown in Fig.\,\ref{perm_schem}.

\begin{figure}[H]
\centering
{%
\resizebox*{10cm}{!}{\includegraphics{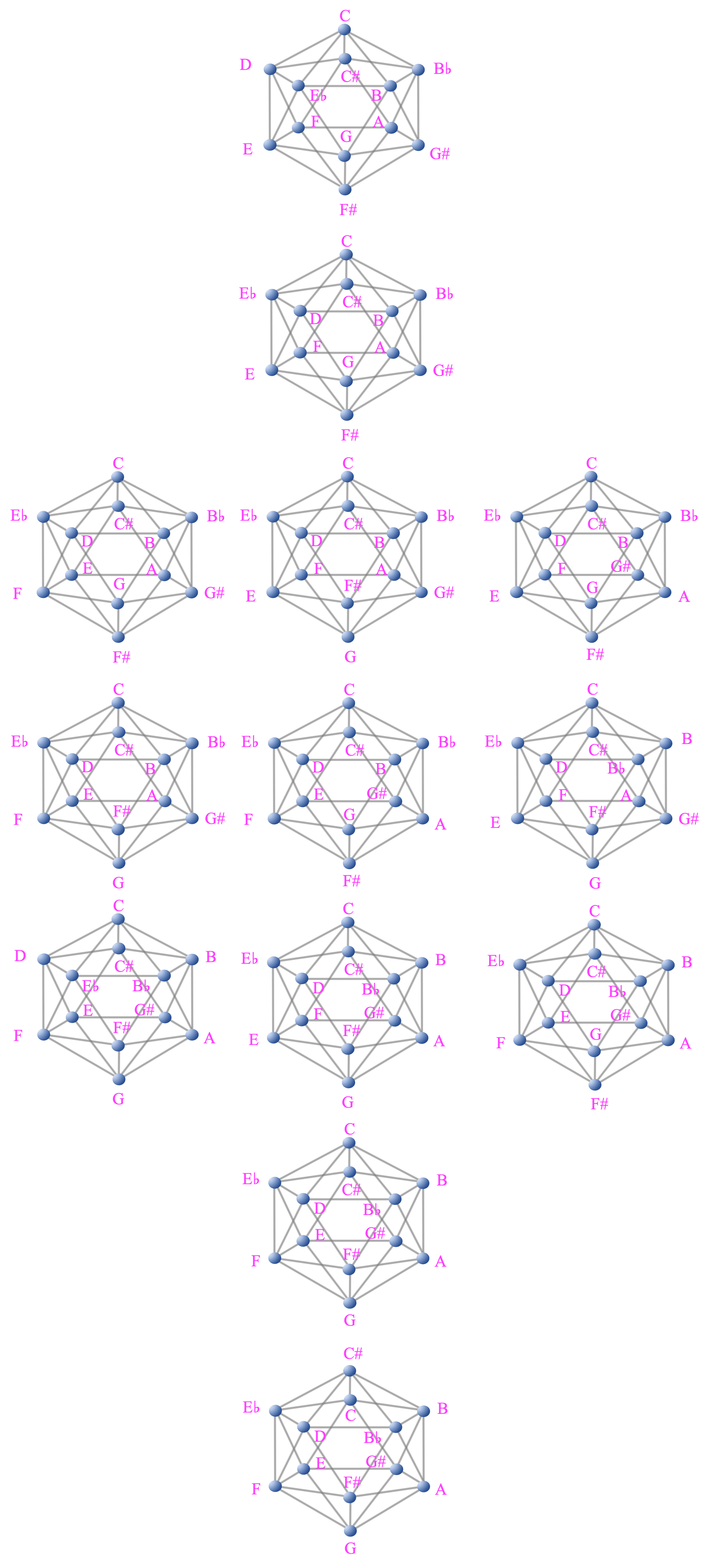}}}\hspace{5pt}
\caption{All the intermediate musical icosahedra from the type 1 to the type 3 corresponding to the diagrams on the left side in Fig.\,\ref{structure_diagram}.} \label{perm_schem_all}
\end{figure}

\begin{figure}[H]
\centering
{%
\resizebox*{14cm}{!}{\includegraphics{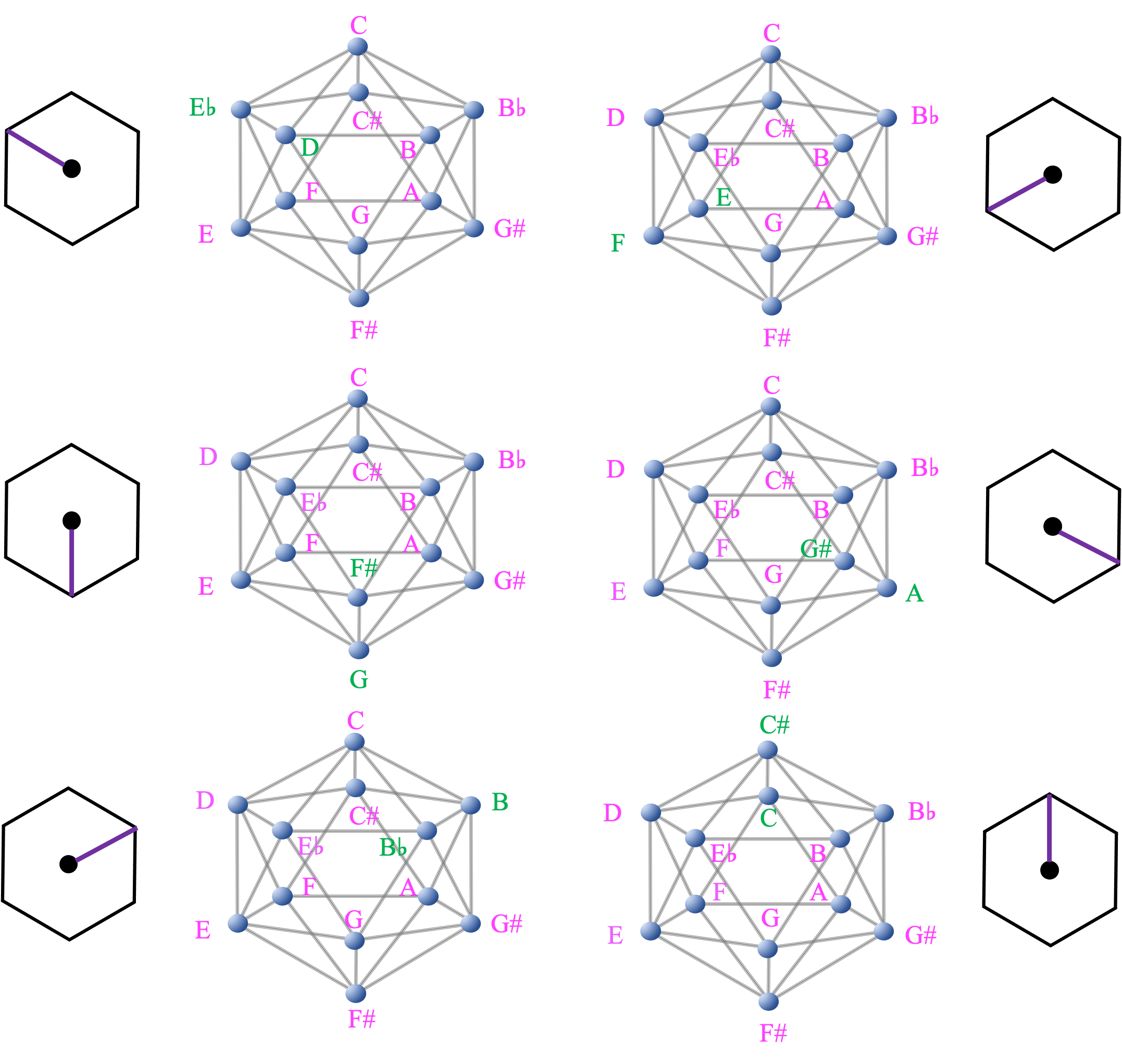}}}\hspace{5pt}
\caption{Inter-permutations that are symmetrically equivalent to the inter-permutation represented by the diagram on the second row in Fig.\,\ref{structure_diagram}.} \label{schem_perm_all}
\end{figure}

In addition, inter-permutations that are symmetrically equivalent to the inter-permutation represented by the diagram on the second row in Fig.\,\ref{structure_diagram} are represented as shown in Fig.\,\ref{schem_perm_all}. There exist 6 inter-permutations that are symmetrically equivalent to the inter-permutation on the third row and the first column. There exist 6 inter-permutations that are symmetrically equivalent to the inter-permutation on the third row and the second column. There exist 3 inter-permutations that are symmetrically equivalent to the inter-permutation on the third row and the third column. There exist 6 inter-permutations that are symmetrically equivalent to the inter-permutation on the fourth row and the first column.There exist 12 inter-permutations that are symmetrically equivalent to the inter-permutation on the fourth row and the second column. There exist 2 inter-permutations that are symmetrically equivalent to the inter-permutation on the third row and the fourth column. The fifth row, the sixth row, the seventh row are equivalent to the third row, the second row, the first row. Note that 1 + 6 + (6 + 6 + 3) + (6 + 12 + 2) + (6 + 6 + 3) + 6 + 1 = 64 = $2^6$.

\begin{figure}[H]
\centering
{%
\resizebox*{14cm}{!}{\includegraphics{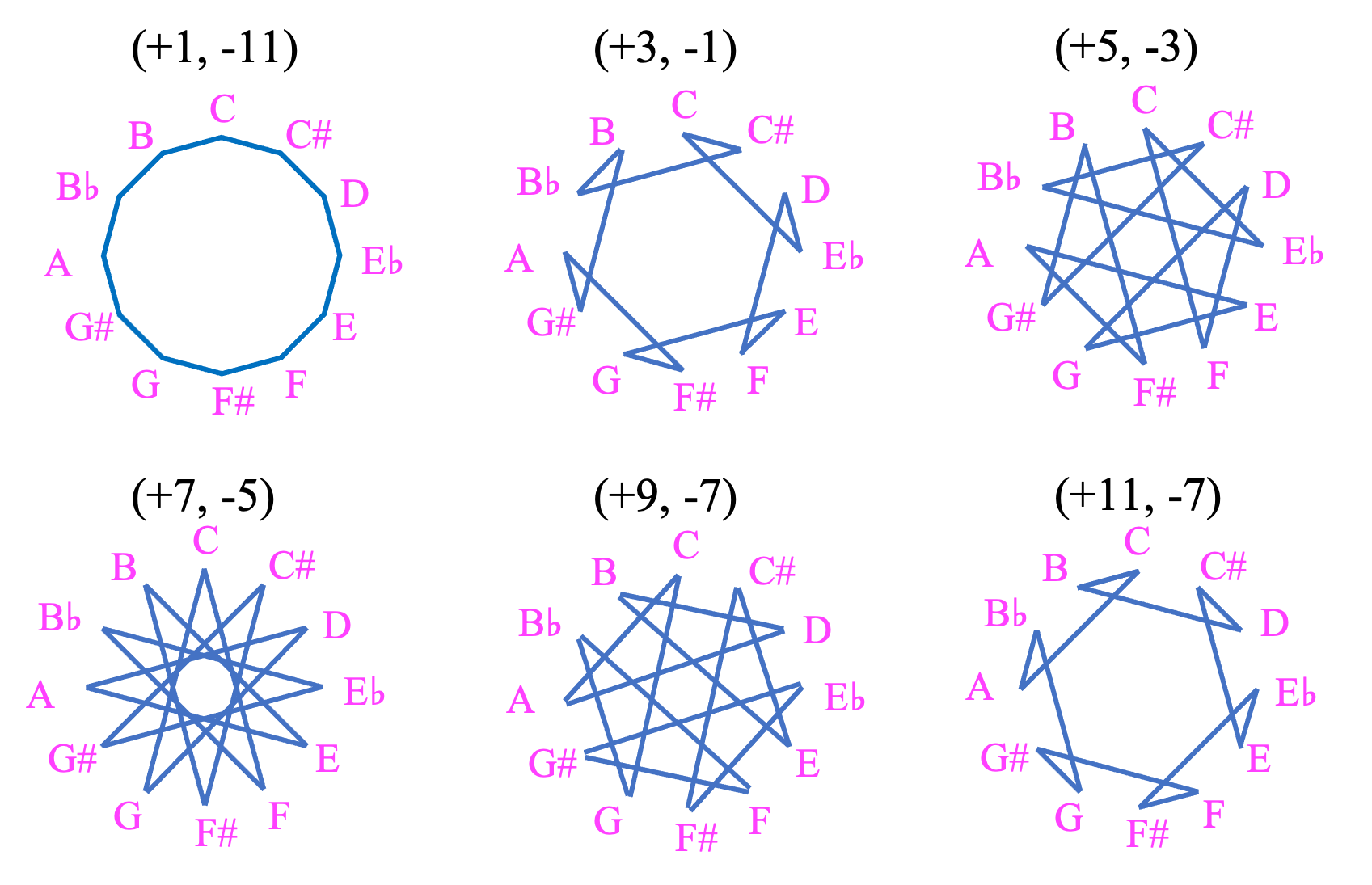}}}\hspace{5pt}
\caption{Schematic representation of the (1, -11) scale, the (3, -1) scale, the (5, -3) scale, the (7, -5) scale, the (9, -7) scale, the (11, -9) scale by the natural arrangement.} \label{12_modi}
\end{figure}

We call the arrangement of 12 tones in the left top figure of Fig.\,\ref{12} natural arrangement. If the natural arrangement is used for the schematic representations of all the kinds of the intermediate musical icosahedra, one has the diagram representing the (1, -11) scale, the (3, -1) scale, the (5, -3) scale, the (7, -5) scale, the (9, -7) scale, the (11, -9) scale as shown in Fig.\,\ref{12_modi}. While the (1, -11) scale and the (7, -5) scale in Fig.\,\ref{12_modi} have the 12-fold symmetry, the other scales do not have the 12-fold symmetry, but the 6-fold symmetry.


\begin{figure}[H]
\centering
{%
\resizebox*{14cm}{!}{\includegraphics{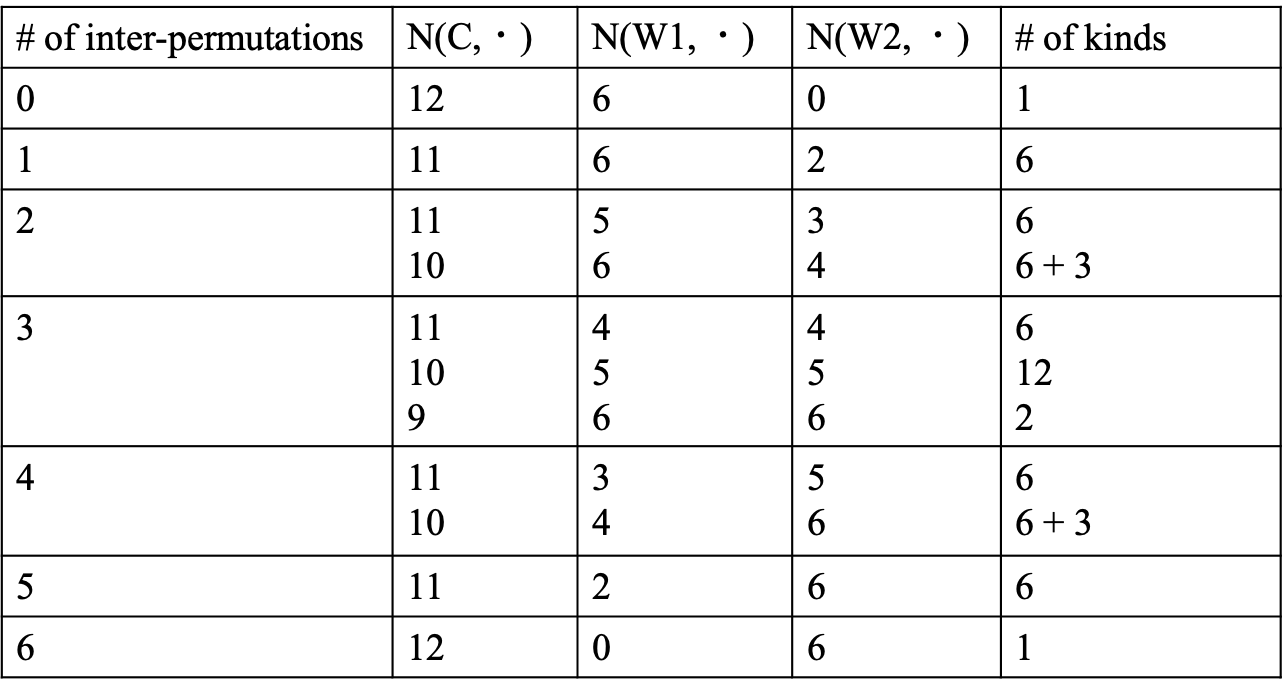}}}\hspace{5pt}
\caption{Structure of the intermediate musical icosahedra from the type 1 to the type 3.} \label{structure_inter}
\end{figure}

The structure of the intermediate musical icosahedra from the type 1 to the type 3 can be summarized in Fig.\,\ref{structure_inter}. The second row shows the type 1. The third row shows musical icosahedra obtained by applying one inter-permutation for the type 1. This case includes one pattern characterized by $N(C, \ \cdot \ )=11$, $N(W1, \  \cdot \ ) = 6$, $N(W2, \  \cdot \ ) = 2$, and including 6 musical icosahedra. The fourth row shows musical icosahedra obtained by applying two inter-permutations for the type 1. This case includes three patterns: one of them is characterized by $N(C, \ \cdot \ )=11$, $N(W1, \ \cdot \ ) = 5$, $N(W2, \ \cdot \ ) = 3$, and includes 6 musical icosahedra, another one is characterized by $N(C, \  \cdot \ )=10$, $N(W1, \ \cdot \ ) = 6$, $N(W2, \  \cdot \ ) = 4$ and includes 6 musical icosahedra, and the other one is also characterized by $N(C, \  \cdot \ )=10$, $N(W1, \ \cdot \ ) = 6$, $N(W2, \ \cdot \ ) = 4$ and includes 3 musical icosahedra. The fifth row shows musical icosahedra obtained by applying three inter-permutations for the type 1. This case includes three patterns: three patterns: one of them is characterized by $N(C, \  \cdot \ )=11$, $N(W1, \  \cdot \ ) = 4$, $N(W2, \  \cdot \ ) = 4$, and includes 6 musical icosahedra, another one is characterized by $N(C, \  \cdot \ )=10$, $N(W1,  \ \cdot \ ) = 5$, $N(W2, \ \cdot \ ) = 5$ and includes 12 musical icosahedra, and the other one is characterized by $N(C, \ \cdot \ )=10$, $N(W1, \  \cdot \ ) = 6$, $N(W2, \  \cdot \ ) = 6$ and includes 2 musical icosahedra. The sixth row shows musical icosahedra obtained by applying four inter-permutations for the type 1. This case is equivalent to the fourth row. The seventh row shows musical icosahedra obtained by applying five inter-permutations for the type 1. This case is equivalent to the third row. The eighth row shows musical icosahedra obtained by applying six inter-permutations for the type 1. This case is equivalent to the second row, and includes the type 3.

Each row of Fig.\,\ref{structure_inter} has the following relations: $2 \times (3 - \sharp \ {\rm of \ inter\mathchar`-permutations}) = N(W_1, \cdot \ ) -N(W_2, \cdot \ )$, $2N(C, \ \cdot \ ) + N(W_1, \ \cdot \ ) + N(W_2, \ \cdot \ ) = 30$, ${}_6{\rm C}_{\rm \sharp \ of \ inter\mathchar`-permutations} = {\rm sum \ of \ \sharp \ of \ kinds }$.

\subsection{Intra-permutation}
In the previous subsection, we introduced the following concepts: the intra-permutation, the neighboring number, the musical invariant, and the intermediate musical icosahedra from the type 1 (type 3) to the type 3 (type 1), from the type 2 (type 4) to the type 4 (type 2), from the type 1' (type 3') to the type 3' (type 1'), from the type 2' (type 4') to the type 4' (type 2'),  from the type ${\rm 1^*}$ (type ${\rm 3^*}$) to the type ${\rm 3^*}$ (type ${\rm 1^*}$), from the type ${\rm 2^*}$ (type ${\rm 4^*}$) to the type ${\rm 4^*}$ (type ${\rm 2^*}$).

In this subsection, we introduce intermediate musical icosahedra from the type $n$ chromatic/whole tone musical icosahedron to the type $n$' Pythagorean/whole tone musical icosahedron by a new kind of permutations, intra-permutation. We call any permutation of 12 tones that fixes the 6 tones set on the regular hexagon in the projection of a given musical icosahedron intra-permutation onto the paper. Namely, the intra-permutations change arrangements of 6 tones set on the vertices of the regular hexagram in the projection of a given musical icosahedron onto the paper. Note that the inter-permutations are permutations that interchange a tone set on the regular hexagon in the projection of a given musical icosahedron onto the paper and a tone set on the regular hexagram in the projection of a given musical icosahedron onto the paper.

We remark that the permutations $C\sharp \leftrightarrow G$, $E\flat \leftrightarrow A$, $F \leftrightarrow B$ for the type n leads to the type n'. Then, we define the intermediate musical icosahedra from the type n to the type n' as musical icosahedra that are obtained by applying some intra-permutations to the type n and that are not the type n or the type n'. Then, there are 718 (=6!-2) intermediate musical icosahedra from the type n to the type n'. Note that the intermediate musical icosahedra from the type 1 to the type 1' include the type 2, the type 2', the type ${\rm 1^*}$, and the type ${\rm 2^*}$ because they are obtained by applying some intra-permutations to the type 1.

\begin{figure}[H]
\centering
{%
\resizebox*{16cm}{!}{\includegraphics{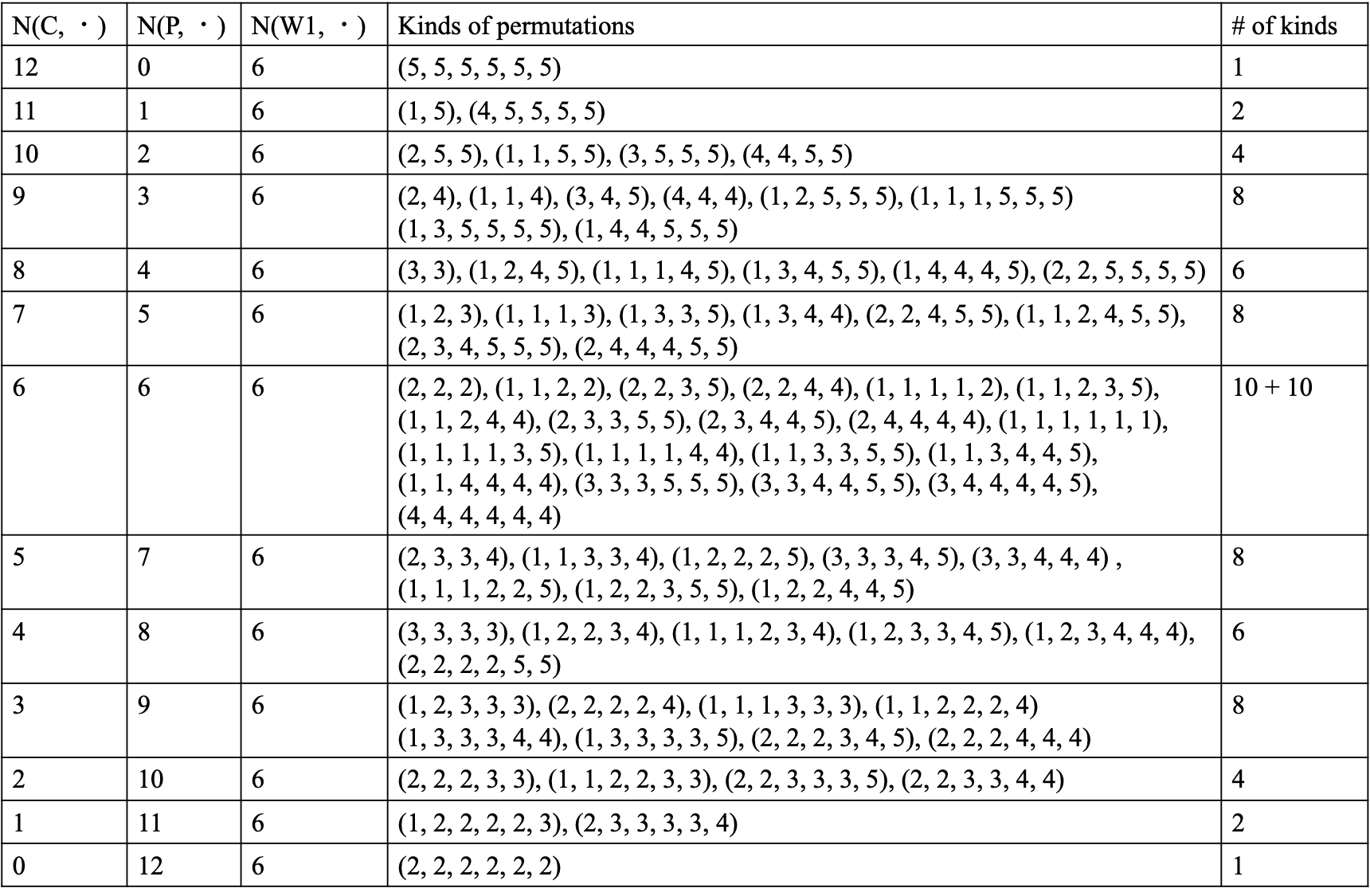}}}\hspace{5pt}
\caption{Structure of the intermediate musical icosahedra from the type 1 to the type 1'.} \label{intra}
\end{figure}

Then, we have the result shown in Fig.\,\ref{intra}. Let me explain a definition used in Fig.\,\ref{intra}. For a natural number less than 6, $n$, natural numbers less than 7, $a_1, \cdots, a_n$, $(a_1, \cdots, a_n)$ is defined as permutations including the following transformations for $b_1, \cdots, b_n$ in $\{C\sharp, E\flat, F, G, A, B\}$, $b_1 \to b_1 + 2a_1$, $\cdots$, $b_n \to b_n + 2b_n$.

There exist 1 kind of the intermediate musical icosahedra from the type 1 to the type 1', (5, 5, 5, 5, 5), satisfies $N(C, \ \cdot \ )=12, \ N(P, \ \cdot \ )=0$. There exist 2 kinds of the intermediate musical icosahedra from the type 1 to the type 1', (1, 5), (4, 5, 5, 5, 5), satisfy $N(C, \ \cdot \ )=11, \  N(P,  \ \cdot \ )=1$. There exist 4 kinds of the intermediate musical icosahedra from the type 1 to the type 1', (2, 5, 5), (1, 1, 5, 5), (3, 5, 5, 5), (4, 4, 5, 5), satisfy $N(C, \ \cdot \ )=10, \  N(P, \ \cdot \ )=2$. There exist 8 kinds of the intermediate musical icosahedra from the type 1 to the type 1', (2, 4), (1, 1, 4), (3, 4, 5), (4, 4, 4), (1, 2, 5, 5, 5), (1, 1, 1, 5, 5, 5), (1, 3, 5, 5, 5, 5), (1, 4, 4, 5, 5, 5), satisfy $N(C, \ \cdot \ )=9, \  N(P, \ \cdot \ )=3$. There exist 6 kinds of the intermediate musical icosahedra from the type 1 to the type 1', (3, 3), (1, 2, 4, 5), (1, 1, 1, 4, 5), (1, 3, 4, 5, 5), (1, 4, 4, 4, 5), (2, 2, 5, 5, 5, 5), satisfy $N(C, \ \cdot \ )=8, \  N(P, \ \cdot \ )=4$. There exist 8 kinds of the intermediate musical icosahedra from the type 1 to the type 1', (1, 2, 3), (1, 1, 1, 3), (1, 3, 3, 5), (1, 3, 4, 4), (2, 2, 4, 5, 5), (1, 1, 2, 4, 5, 5), (2, 3, 4, 5, 5, 5), (2, 4, 4, 4, 5, 5), satisfy $N(C, \ \cdot \ )=7, \ N(P, \ \cdot \ )=5$. There exist 20 kinds of the intermediate musical icosahedra from the type 1 to the type 1', (2, 2, 2), (1, 1, 2, 2), (2, 2, 3, 5), (2, 2, 4, 4), (1, 1, 1, 1, 2), (1, 1, 2, 3, 5), (1, 1, 2, 4, 4), (2, 3, 3, 5, 5), (2, 3, 4, 4, 5), (2, 4, 4, 4, 4), (1, 1, 1, 1, 1, 1), (1, 1, 1, 1, 3, 5), (1, 1, 1, 1, 4, 4), (1, 1, 3, 3, 5, 5), (1, 1, 3, 4, 4, 5), (1, 1, 4, 4, 4, 4), (3, 3, 3, 5, 5, 5), (3, 3, 4, 4, 5, 5), (3, 4, 4, 4, 4, 5), (4, 4, 4, 4, 4, 4), satisfy $N(C, \ \cdot \ )=6, \ N(P, \ \cdot \ )=6$. There exist 8 kinds of the intermediate musical icosahedra from the type 1 to the type 1', (2, 3, 3, 4), (1, 1, 3, 3, 4), (1, 2, 2, 2, 5), (3, 3, 3, 4, 5), (3, 3, 4, 4, 4), (1, 1, 1, 2, 2, 5), (1, 2, 2, 3, 5, 5), (1, 2, 2, 4, 4, 5), satisfy $N(C, \ \cdot \ )=5, \ N(P, \ \cdot \ )=7$. There exist 6 kinds of the intermediate musical icosahedra from the type 1 to the type 1', (3, 3, 3, 3), (1, 2, 2, 3, 4), (1, 1, 1, 2, 3, 4), (1, 2, 3, 3, 4, 5), (1, 2, 3, 4, 4, 4), (2, 2, 2, 2, 5, 5), satisfy $N(C, \ \cdot \ )=4, \ N(P, \ \cdot \ )=8$. There exist 8 kinds of the intermediate musical icosahedra from the type 1 to the type 1', (1, 2, 3, 3, 3), (2, 2, 2, 2, 4), (1, 1, 1, 3, 3, 3), (1, 1, 2, 2, 2, 4), (1, 3, 3, 3, 4, 4), (1, 3, 3, 3, 3, 5), (2, 2, 2, 3, 4, 5), (2, 2, 2, 4, 4, 4), satisfy $N(C, \ \cdot \ )=3, \ N(P, \ \cdot \ )=9$. There exist 4 kinds of the intermediate musical icosahedra from the type 1 to the type 1', (2, 2, 2, 3, 3), (1, 1, 2, 2, 3, 3), (2, 2, 3, 3, 3, 5), (2, 2, 3, 3, 4, 4), satisfy $N(C, \ \cdot \ )=2, \ N(P, \ \cdot \ )=10$. There exist 2 kinds of the intermediate musical icosahedra from the type 1 to the type 1', (1, 2, 2, 2, 2, 3), (2, 3, 3, 3, 3, 4), satisfy $N(C, \ \cdot \ )=1, \  N(P, \ \cdot \ )=11$. There exist 1 kind of the intermediate musical icosahedra from the type 1 to the type 1', (2, 2, 2, 2, 2, 2), satisfies $N(C, \ \cdot \ )=0, \ N(P, \ \cdot \ )=12$. All the kinds of the intermediate musical icosahedra from the type 1 to the type 1' satisfy $N(W1, \ \cdot \ )=6$.

\begin{figure}[H]
\centering
{%
\resizebox*{14cm}{!}{\includegraphics{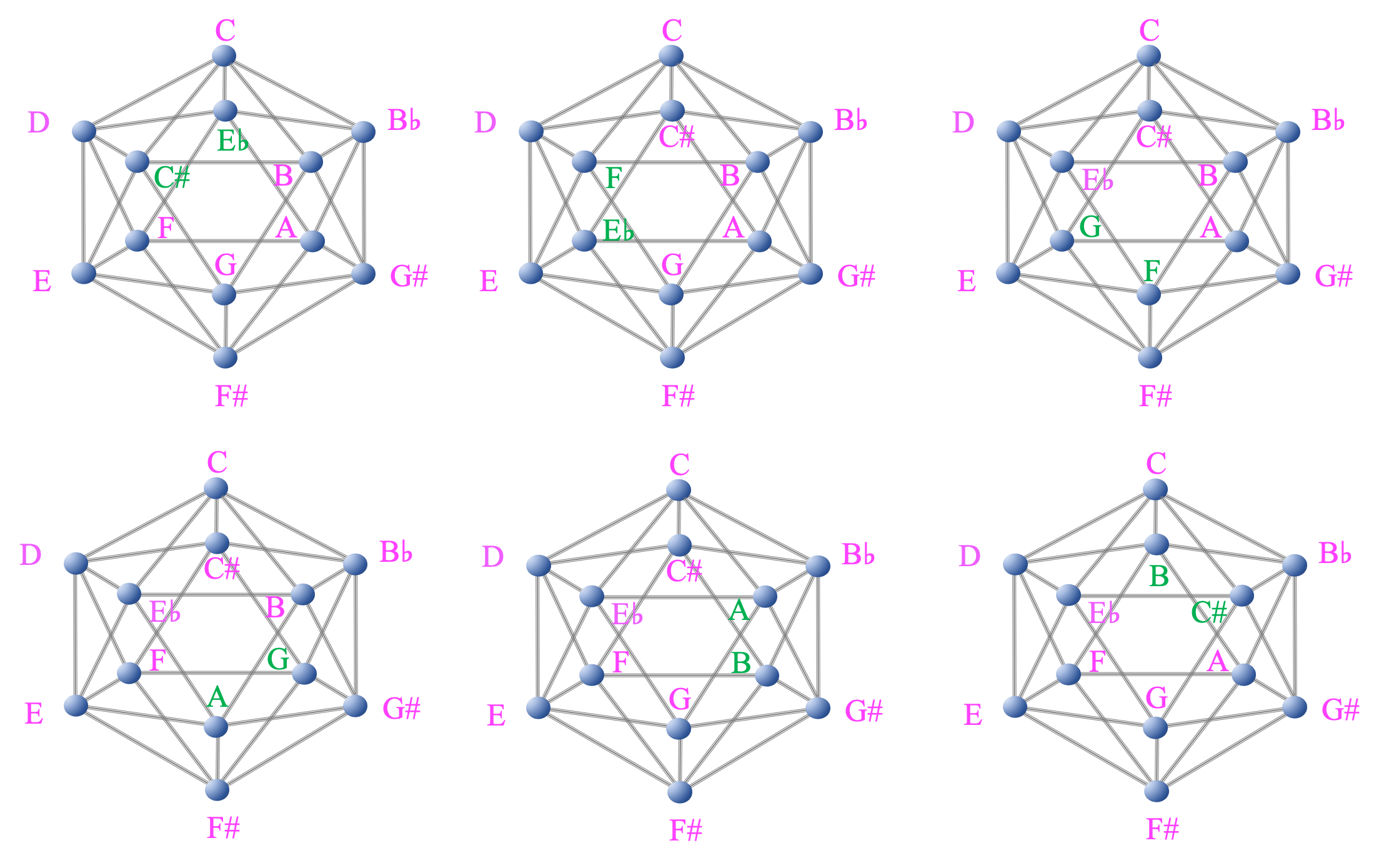}}}\hspace{5pt}
\caption{Musical icosahedra included in the (1, 5).} \label{15}
\end{figure}

Each kind in Fig.\,\ref{intra} corresponds to one or more musical icosahedra. For example, (1, 5) includes the intermediate musical icosahedra from the type 1 to the type 1' shown in Fig.\,\ref{15}.

We remark that $N(C, \ \cdot \ ) + N(P, \ \cdot \ )=12$, $N(W1, \cdot \ )=6$ for the type 1 and the type 1' and all the intermediate musical icosahedra from the type 1 to the type 1'. Note that for a tone $X$, $X\pm 1 = (X+6) \pm7$, and each of $C$, $D$, $E$ is set oppositely to each of $F\sharp$, $G\sharp$, $B\flat$ for the type 1 and the type 1' and all the intermediate musical icosahedra from the type 1 to the type 1'.

\newpage
\section{Melakarta Raga and Musical Icosahedron}
In this section, we show how the intra-permutations, the inter-permutations, the chromatic/whole tone musical icosahedra, and the major scale lead to the Melakartra raga.

\subsection{Prmutation-extension}
In this subsection, we define permutation-extension. The permutation-extension of a given scale based on $C$ is obtained by the following procedure. First, color the vertices corresponding to tones constructing a given scale based on $C$ on a given musical icosahedron. Second, color vertices on the musical icosahedra obtained by applying some inter-permutations to the given musical icosahedron, that are set in the same position as the colored vertices on the given musical icosahedron are set. An ascending tone sequence based on $C$ can be uniquely obtained by the colored vertices on each of those musical icosahedra. We define permutation-extension of the given scale by the given musical icosahedron, by a set of those obtained ascending tone sequences.

We introduce the following notations, $E(A, X)$ and $E(A, X_1, \cdots, X_n)$. $E(A, X)$ means a permutation-extension of $A$ by a musical icosahedron $X$, and $E(A, X_1, \cdots, X_n) = \cup_{i=i}^n E(A, X_i)$.

\subsection{Melakarta Raga as Permutation-extension of Major Scale}
\begin{figure}[H]
\centering
{%
\resizebox*{14cm}{!}{\includegraphics{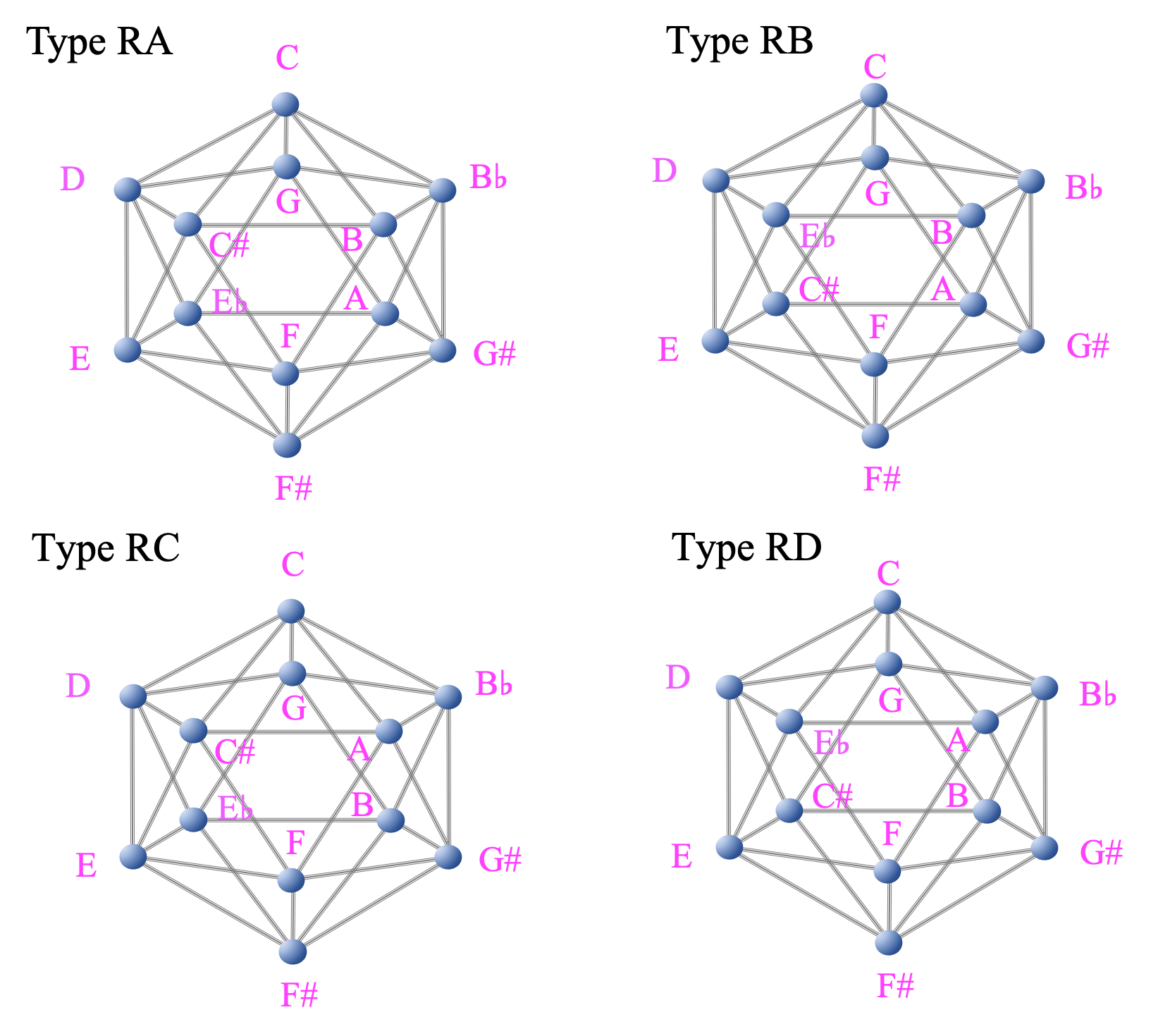}}}\hspace{5pt}
\caption{Four types of the Melakarta raga musical icosahedra.} \label{Raga}
\end{figure}

In this subsection, we define four types of musical icosahedra: the Melakarta raga musical icosahedra (Fig.\,\ref{Raga}). All the types are included in the intermediate musical icosahedra from the type 1 to the type 1'. The type RA is included in the (3, 5, 5, 5), the type RB in the (3, 4, 5), the type RC in the (1, 3, 5, 5, 5, 5), the type RD in the (1, 3, 4, 5, 5). Therefore, all of them, and the type 1, and the type 1' satisfy the following condition of musical invariants: $N(C, P, 1, 1, \  \cdot \ ) = 15$, $N(W1, \  \cdot \ )=6$.

\begin{figure}[H]
\centering
{%
\resizebox*{12cm}{!}{\includegraphics{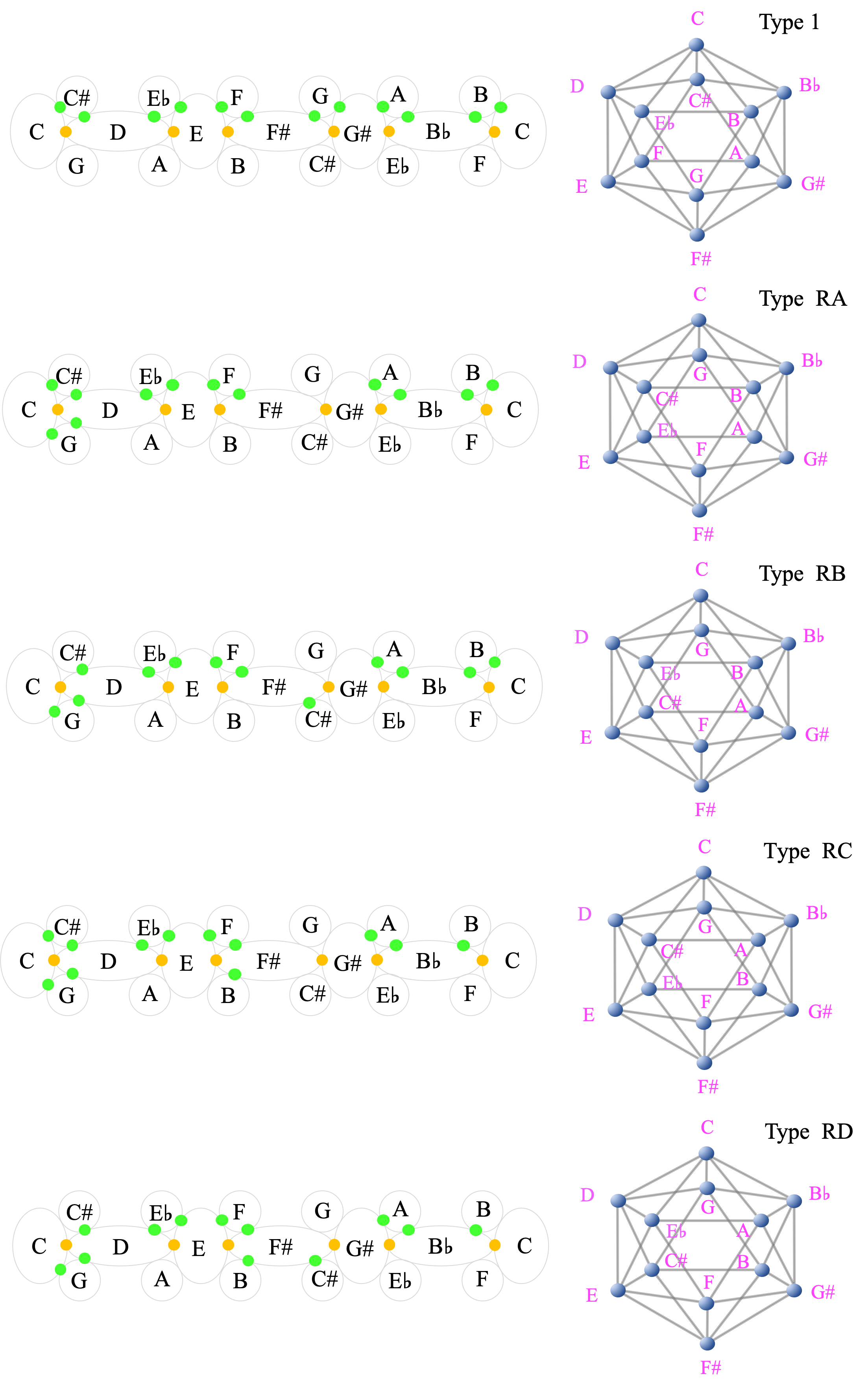}}}\hspace{5pt}
\caption{A visualization of $N(C, P, 1, 1, \  \cdot \ )$ and $N(W1, \ \cdot \ )$ for the type 1, the type 1', and all the types of the Melakarta raga musical icosahedra. The number of the green points equals to 12 and the orange points to 6 for all the musical icosahedron shown here.} \label{index_graph_1}
\end{figure}

This can be schematically represented by using a diagram shown in Fig.\,\ref{index_graph_ex_1}. We show the neighboring property of the chromatic scale, Pythagorean chain and whole tone scale including $C$ in the type 1 and the four types of the Melakarta raga musical icosahedra in Fig.\,\ref{index_graph_1}. The green points are set if the two tones constructing the intersection neighbor each other in a given musical icosahedron and the intersection is made by a circle and an ellipse. The orange points are set if the two tones constructing the intersection neighbor each other in a given musical icosahedron and the intersection is made by ellipses. Therefore, the number of the green points represents $N(C, P, 1, 1, \  \cdot \ )$ and the number of the green points represents $N(W, \ \cdot \ )$. For all the musical icosahedra shown in Fig.\,\ref{index_graph_ex_1}, the number of the green points equals to 12 and the number of the orange points equals to 6. This means that for the type 1, the type 1', and all the types of the Melakarta raga musical icosahedra, $N(C, P, 1, 1, \  \cdot \ ) = 15$, $N(W1, \ \cdot \ )=6$.

\begin{figure}[H]
\centering
{%
\resizebox*{14cm}{!}{\includegraphics{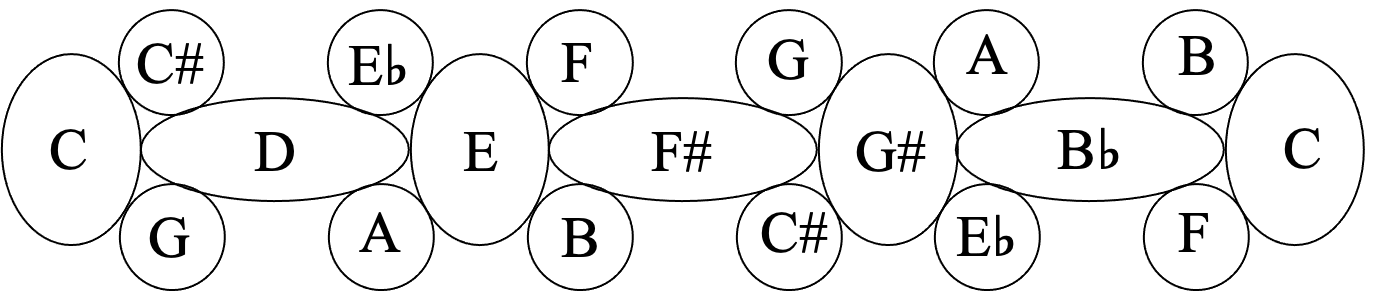}}}\hspace{5pt}
\caption{A diagram for visualization of $N(C, P, 1, 1, \ \cdot \ )$ and $N(W1, \ \cdot \ )$. The intersections of ellipses in the diagram represent the neighboring property of the whole tone scale including $C$ and the intersections of an ellipse and a circle in the diagram represent the neighboring property of the chromatic scale or the Pythagorean chain.} \label{index_graph_ex_1}
\caption{A diagram for visualization of $N(C, P, 1, 1, \ \cdot \ )$ and $N(W1, \ cdot \ )$. The intersections of ellipses in the diagram represent the neighboring property of the whole tone scale including $C$ and the intersections of an ellipse and a circle in the diagram represent the neighboring property of the chromatic scale or the Pythagorean chain.} \label{index_graph_ex_1}
\end{figure}

Then, we consider all the musical icosahedra obtained by performing the iner-permutations to all the types of the Melakarta raga musical icosahedra. We introduce the following four deformed chromatic scales and whole tone scales.
\\
\\
\indent
Chromatic A scale: $C$, $G$, $D$, $C\sharp$, $E$, $E\flat$, $F\sharp$, $F$, $G\sharp$, $A$, $B\flat$, $B$.

Chromatic B scale: $C$, $G$, $D$, $E\flat$, $E$, $C\sharp$, $F\sharp$, $F$, $G\sharp$, $A$, $B\flat$, $B$.

Chromatic C scale: $C$, $G$, $D$, $C\sharp$, $E$, $E\flat$, $F\sharp$, $F$, $G\sharp$, $B$, $B\flat$, $A$.

Chromatic D scale: $C$, $G$, $D$, $E\flat$, $E$, $C\sharp$, $F\sharp$, $F$, $G\sharp$, $B$, $B\flat$, $A$.

Whole tone A2 scale: $G$, $C\sharp$, $E\flat$, $F$, $A$, $B$.

Whole tone B2 scale: $G$, $E\flat$, $C\sharp$, $F$, $A$, $B$.

Whole tone C2 scale: $G$, $C\sharp$, $E\flat$, $F$, $B$, $A$.

\begin{figure}[H]
\centering
{%
\resizebox*{14cm}{!}{\includegraphics{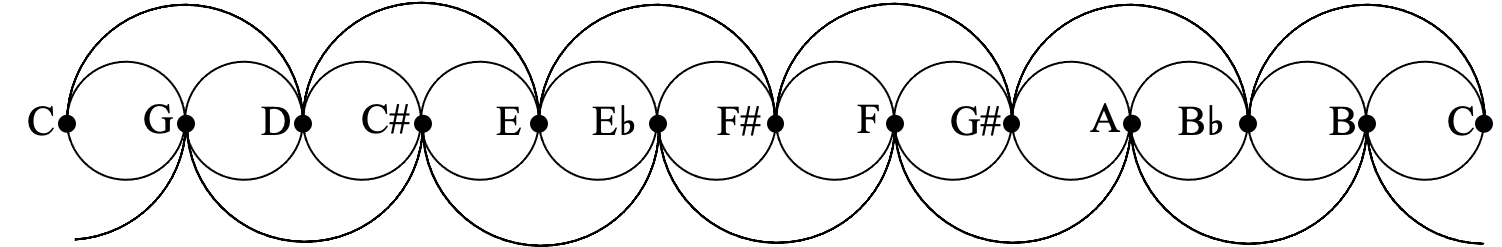}}}\hspace{5pt}
\caption{A diagram for visualization of $N(CA, W1, Wn2, 1, 1/2, 1/2, \ \cdot \ )$.} \label{index_graph_ex_2}
\end{figure}

Whole tone D2 scale: $G$, $E\flat$, $C\sharp$, $F$, $B$, $A$.
\\
\\
\indent
For n=A, B, C, D, $N(Cn, W1, Wn2, 1, 1/2, 1/2, {\rm type \ Rn})=15$ where $Cn$ means the chromatic $n$ scale and $Wn2$ means the whole tone n2 scale. One can visualize $N(CA, W1, WA2, 1, 1/2, 1/2, {\rm type \ RA})$ by a diagram shown in Fig.\ref{index_graph_ex_2}. We call this diagram permutation-circle diagram. In the same manner, one can visualize $N(CB, W1, WB2, 1, 1/2, 1/2, {\rm type \ RB})$, $N(CC, W1, WC2, 1, 1/2, 1/2,{\rm type \ RC})$, and $N(CD, W1, WD2, 1, 1/2, 1/2, {\rm type \ RD})$.

\begin{figure}[H]
\centering
{%
\resizebox*{12cm}{!}{\includegraphics{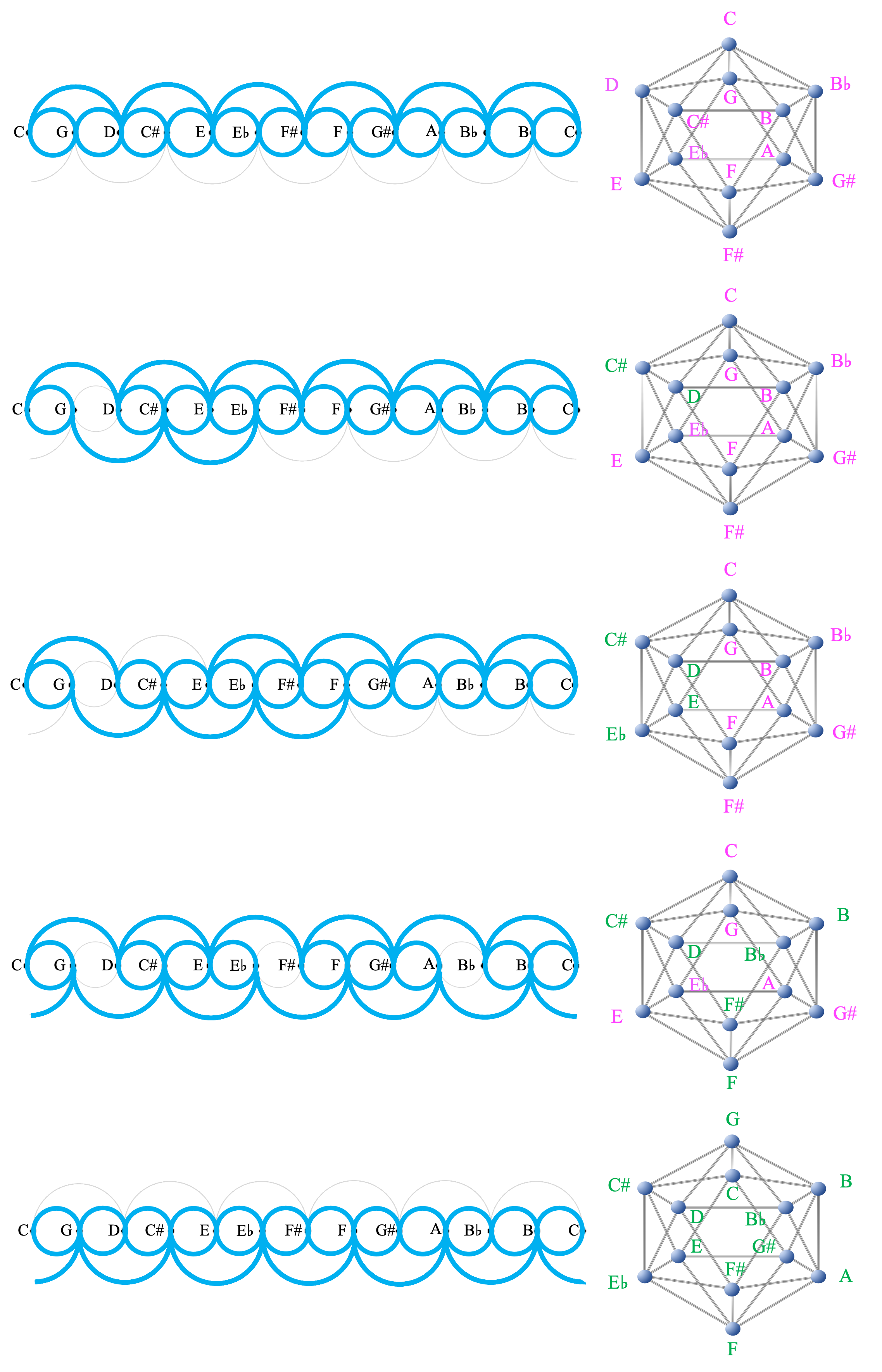}}}\hspace{5pt}
\caption{A visualization of $N(CA, W1, Wn2, 1, 1/2, 1/2, \ \cdot \ )$ for the type RA and some musical icosahedra obtained by performing the iner-permutations to the type RA. For all the musical icosahedra, the number of circles is 15.} \label{index_graph_2}
\end{figure}

We show $N(CA, W1, WA2, 1, 1/2, 1/2, \ \cdot \ )$ for the type RA and some musical icosahedra obtained by performing the inter-permutations to the type RA by the diagram of Fig.\,\ref{index_graph_2}. We color the diagram by the following procedure. If $C$ and $D$ are connected in a given musical icosahedron, the half-circle including $C$ and $D$ is colored. The half-circle including $D$ and $E$, or $E$ and $F\sharp$, or $F\sharp$ and $G\sharp$, or $G\sharp$ and $B\flat$, or $B\flat$ and $C$, or $G$ and $C\sharp$, or $C\sharp$ and $E\flat$, or $E\flat$ and $F$, or $F$ and $A$, or $A$ and $B$, are colored in the same manner. If $B$ and $G$ are connected in a given musical icosahedron, the quarter-circles including $B$ or $G$ are colored. If $C$ and $G$ are connected in a given musical icosahedron, the circle including $C$ and $G$ is colored. The circle including $G$ and $D$, or $D$ and $C\sharp$, or $C\sharp$ and $E$, or $E$ and $E\flat$, or $E\flat$ and $F\sharp$, or $F\sharp$ and $F$, or $F$ and $G\sharp$, or $G\sharp$ and $A$, or $A$ and $B\flat$, or $B\flat$ and $B$, or $B$ and $C$, is colored in the same manner.

Because the number of circles\footnote{We count the number of circles in the following manner: two half-circles equal to a circle and two quarter-circles equal to a half-circle.} for all the diagram in Fig.\,\ref{index_graph_2} equals to 15, and then, $N(CA, W1, Wn2, 1, 1/2, 1/2, \  \cdot \ )=15$ for all the musical icosahedra shown in Fig.\,\ref{index_graph_2}.

\newpage
Then, we obtain the following theorem.
\\
\\
\indent
{\bf Theorem of Melakarta Raga and Permutation-Extension}

\begin{figure}[H]
\centering
{%
\resizebox*{14cm}{!}{\includegraphics{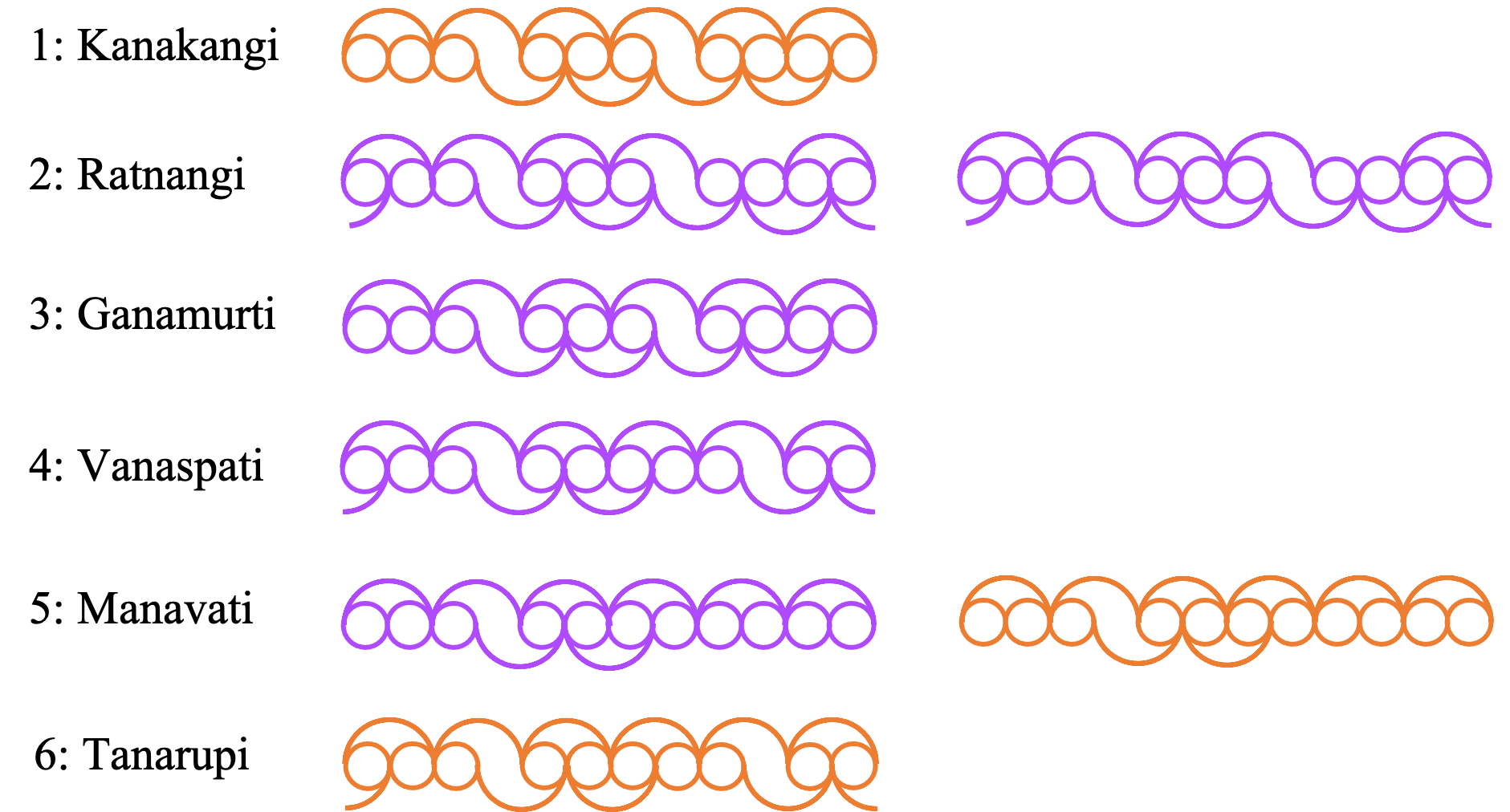}}}\hspace{5pt}
\caption{Representation of Kanakangi, Ratnangi, Ganamurti, Vanaspati, Manavati, Tanarupi by the circle-permutation diagram.} \label{Raga1}
\end{figure}

\begin{figure}[H]
\centering
{%
\resizebox*{14cm}{!}{\includegraphics{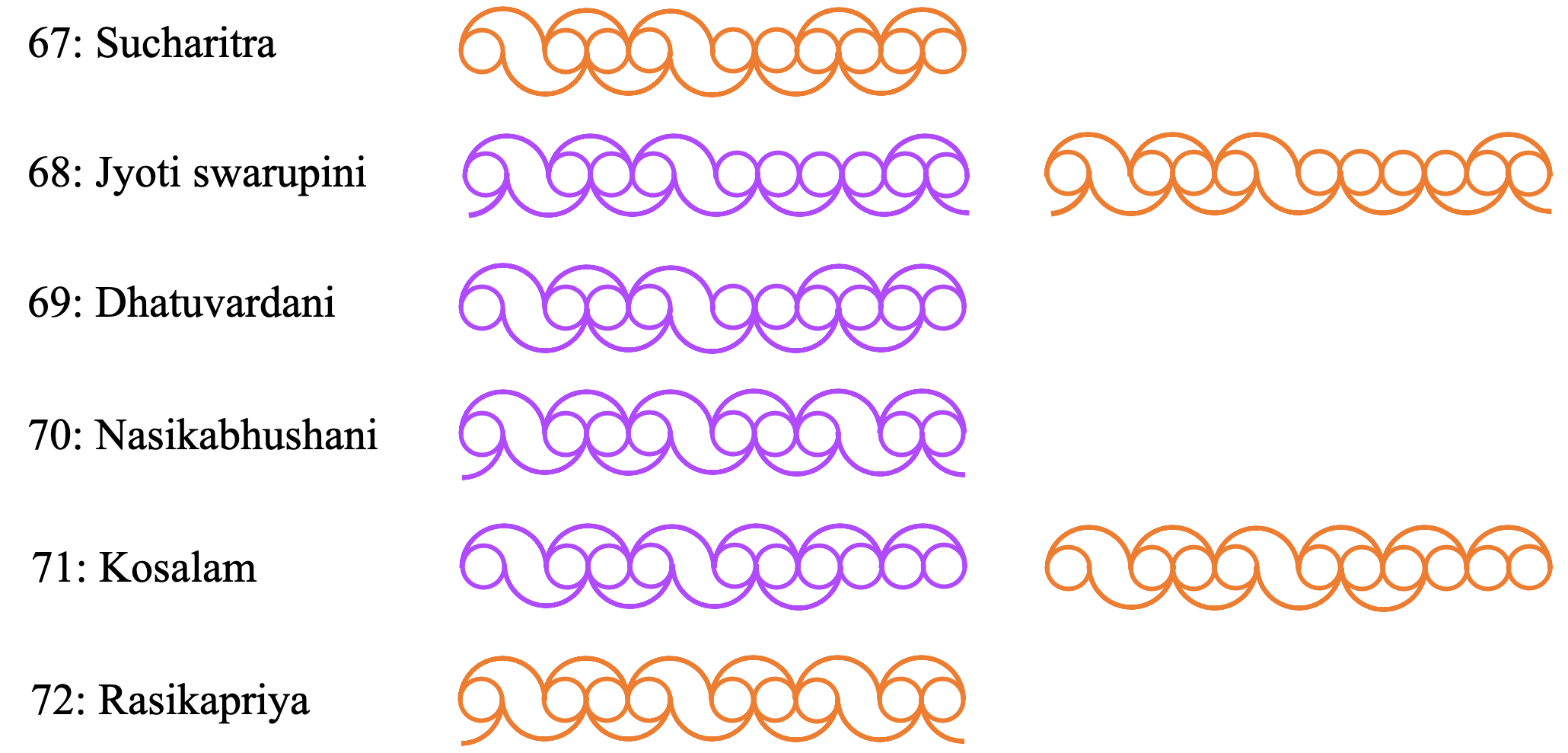}}}\hspace{5pt}
\caption{Representation of Sucharitra, Jyoti swarupini, Dhatuvardani, Nasikabhushani, Kosalam, Rasikapriya by the circle-permutation diagram.} \label{Raga12}
\end{figure}

$E({\rm C \ major \  scale, type RA, type RB, type RC, type RD})$ is a set of all the scales included in the Melakarta raga.
\\
\\
\indent
In Fig.\,\ref{Raga1} - Fig,\,\ref{Raga12}, we show how each scale of Melakarta raga can be characterized by the permutation-extension by using the permutation-circle digram shown in Fig.\,\ref{index_graph_2}. We use sky-blue lines for the permutation-extensions for the type RA, purple lines for the permutation-extensions for the type RB, yellow-green lines for the permutation-extensions for the type RC, orange lines for the permutation-extensions for the type RD.

\begin{figure}[H]
\centering
{%
\resizebox*{14cm}{!}{\includegraphics{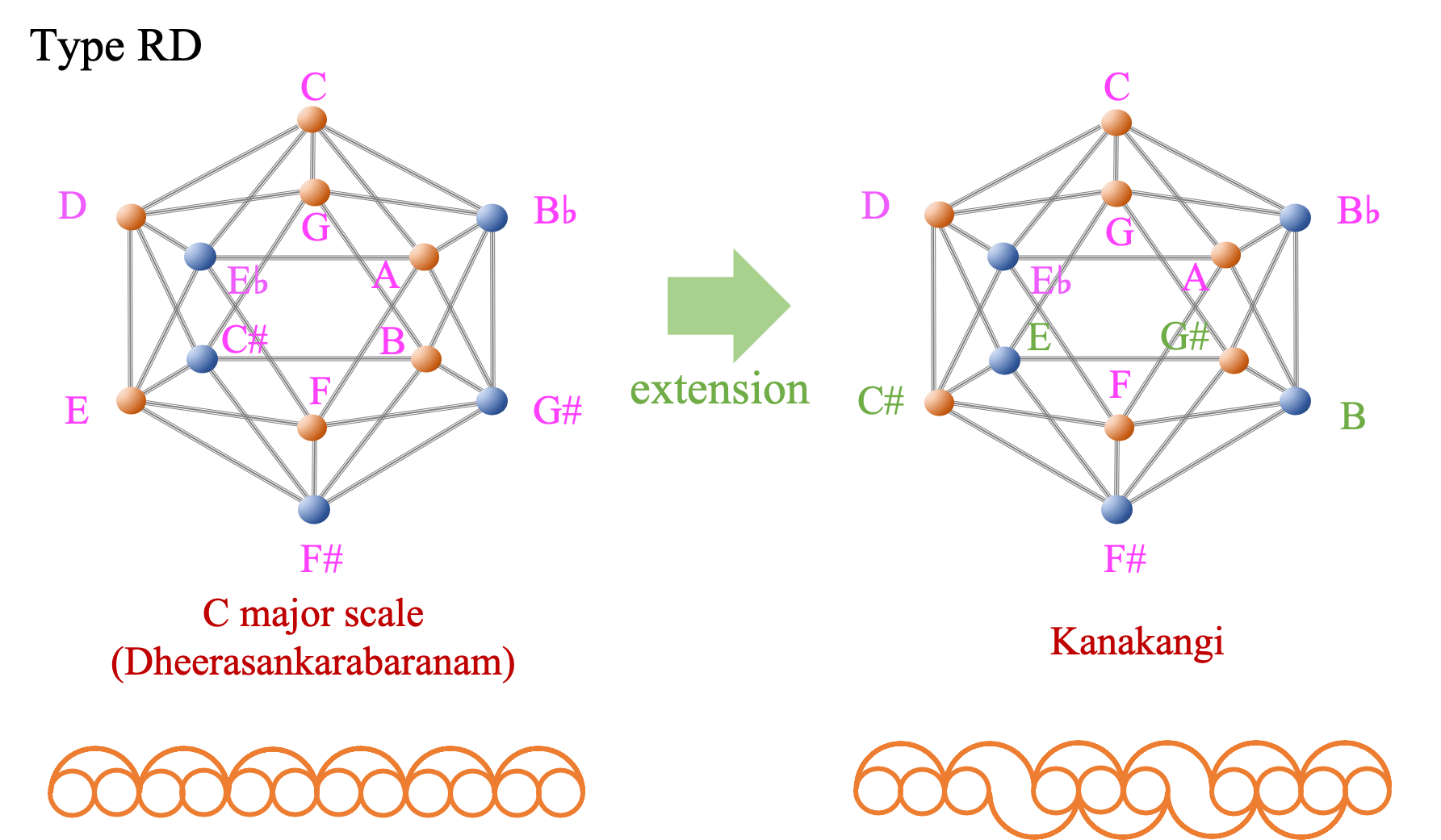}}}\hspace{5pt}
\caption{An example of a scale included in the permutation-extension of the $C$ major scale: Kanakangi
.} \label{extension_ex}
\end{figure}

Also, Fig.\,\ref{extension_ex} shows how Kanakangi is derived by the $C$ major scale.

\begin{figure}[H]
\centering
{%
\resizebox*{14cm}{!}{\includegraphics{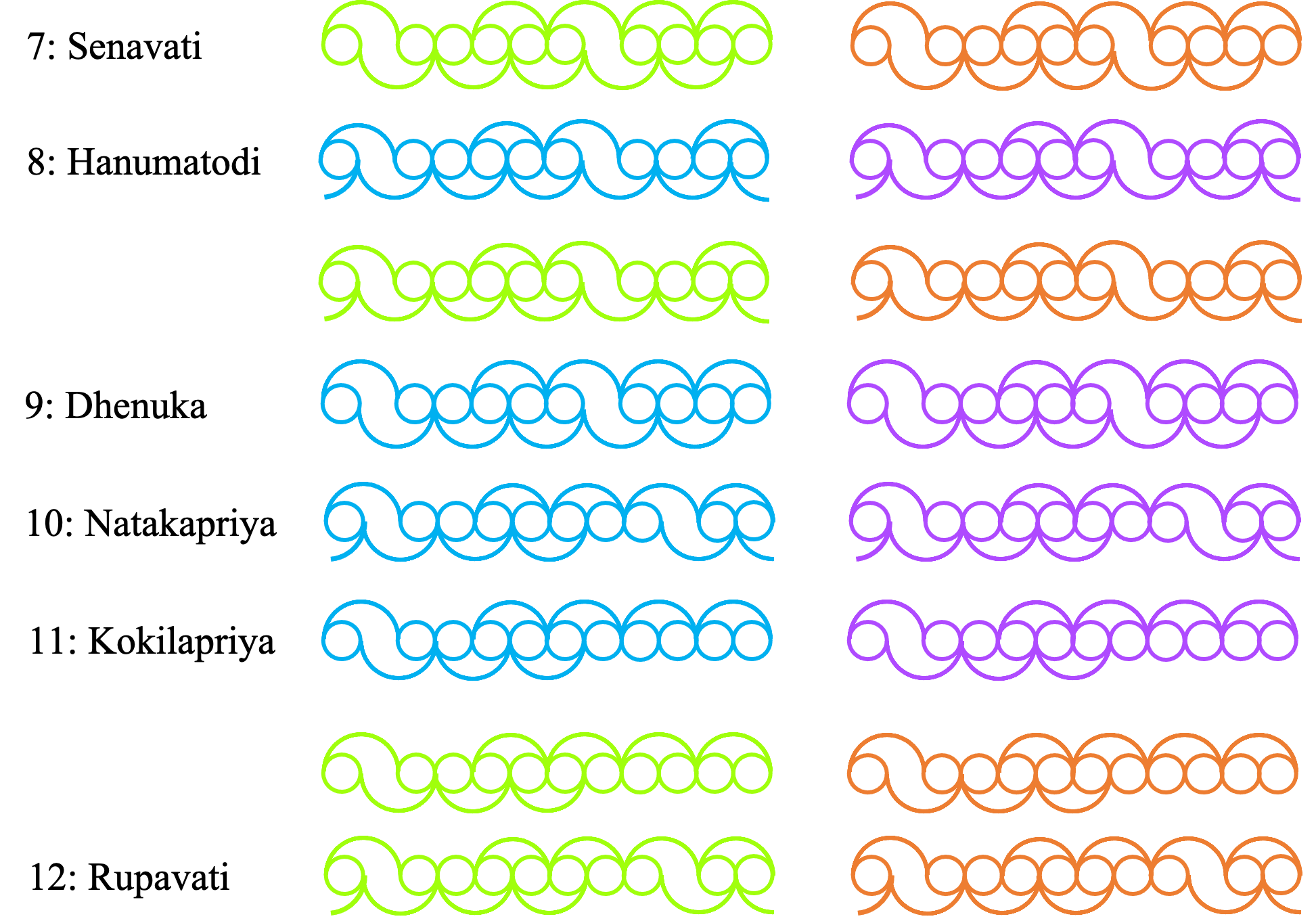}}}\hspace{5pt}
\caption{Representation of Senavati, Hanumatodi, Dhenuka, Natakapriya, Kokilapriya, Rupavati by the circle-permutation diagram.} \label{Raga2}
\end{figure}

\begin{figure}[H]
\centering
{%
\resizebox*{14cm}{!}{\includegraphics{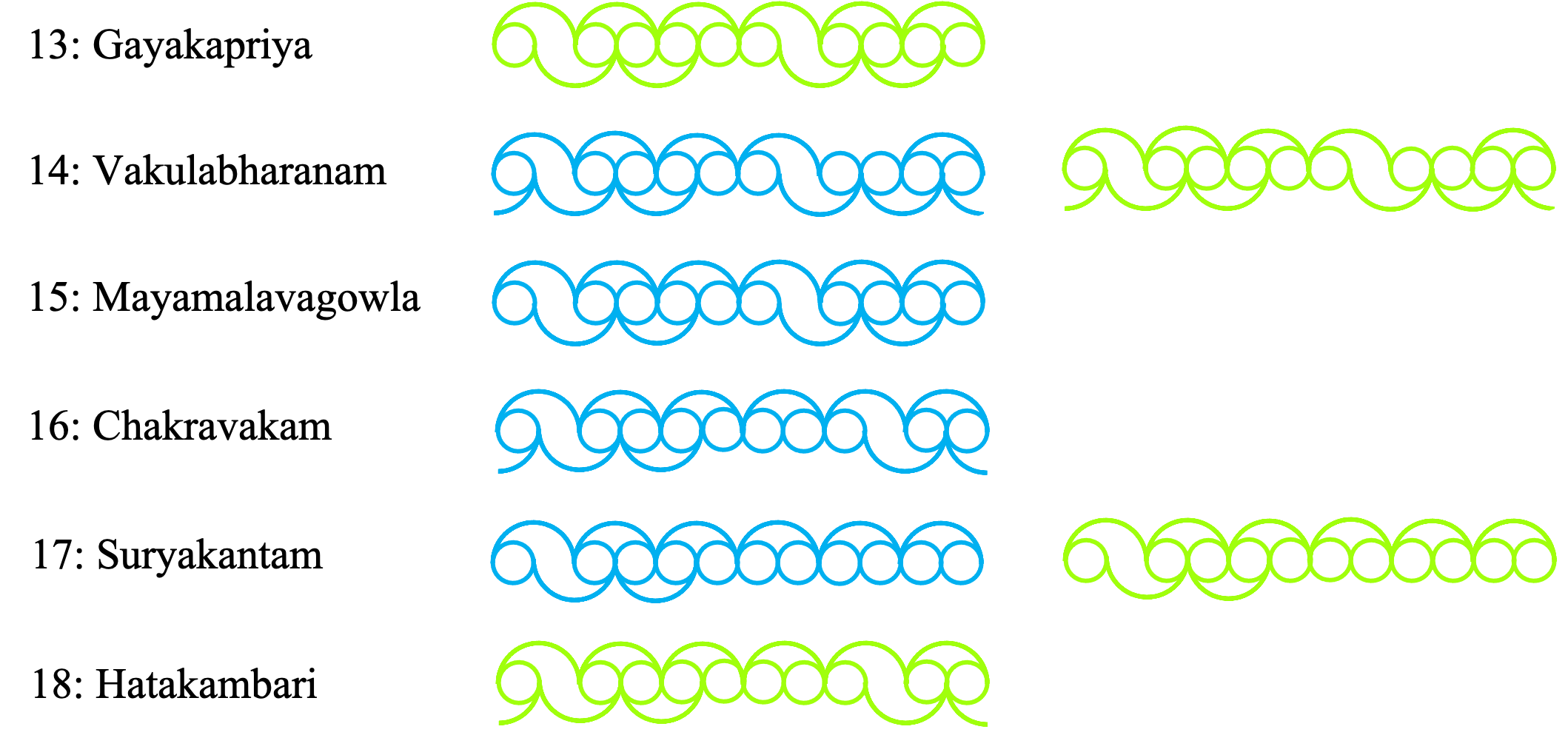}}}\hspace{5pt}
\caption{Representation of Gayakapriya, Vakulabharanam, Mayamalavagowla, Chakravakam, Suryakantam, Hatakambari by the circle-permutation diagram.} \label{Raga3}
\end{figure}

\begin{figure}[H]
\centering
{%
\resizebox*{14cm}{!}{\includegraphics{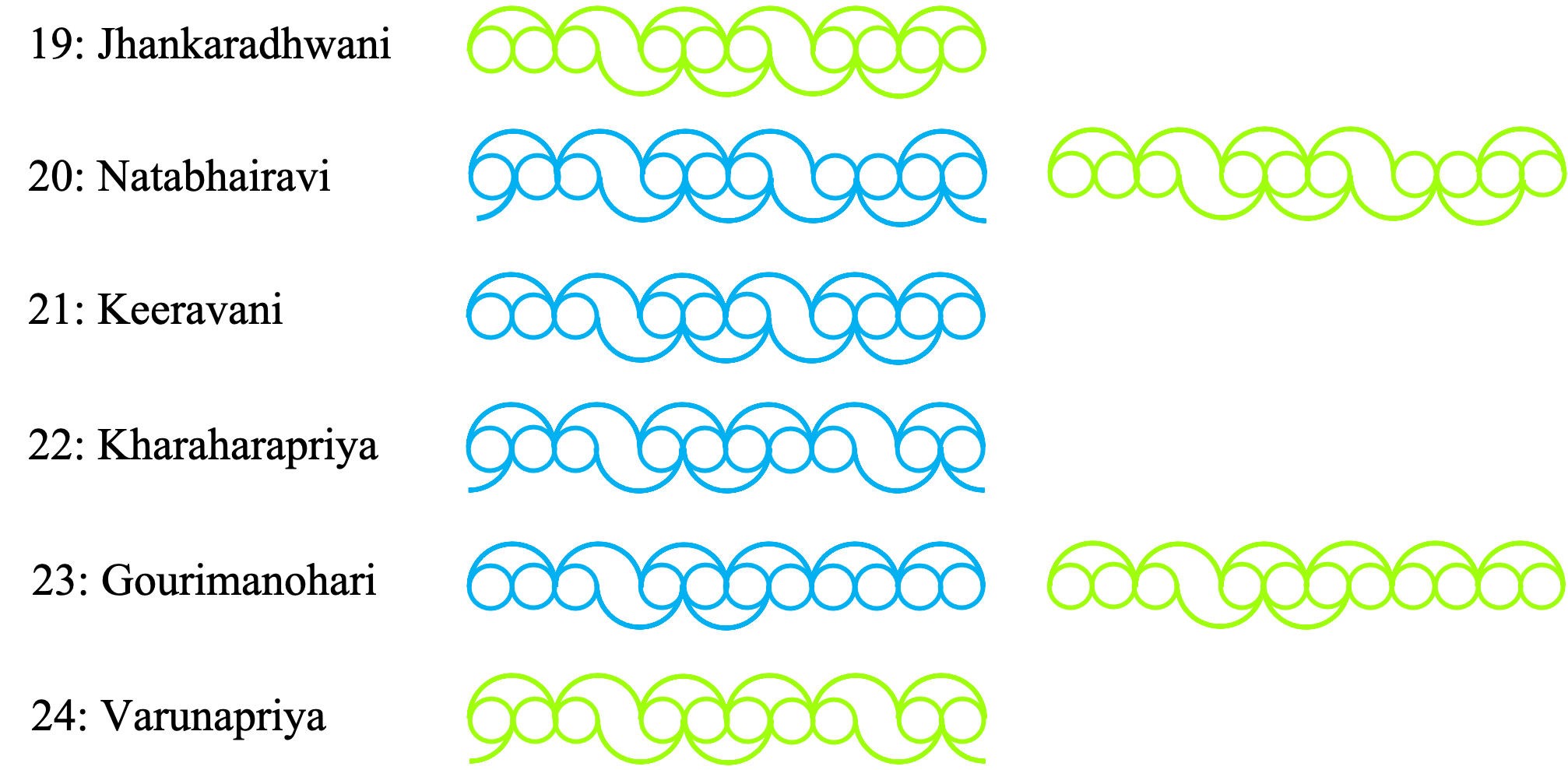}}}\hspace{5pt}
\caption{Representation of Jhankaradhwani, Natabhairavi, Keeravani, Kharaharapriya, Gourimanohari, Varunapriya by the circle-permutation diagram.} \label{Raga4}
\end{figure}

\begin{figure}[H]
\centering
{%
\resizebox*{14cm}{!}{\includegraphics{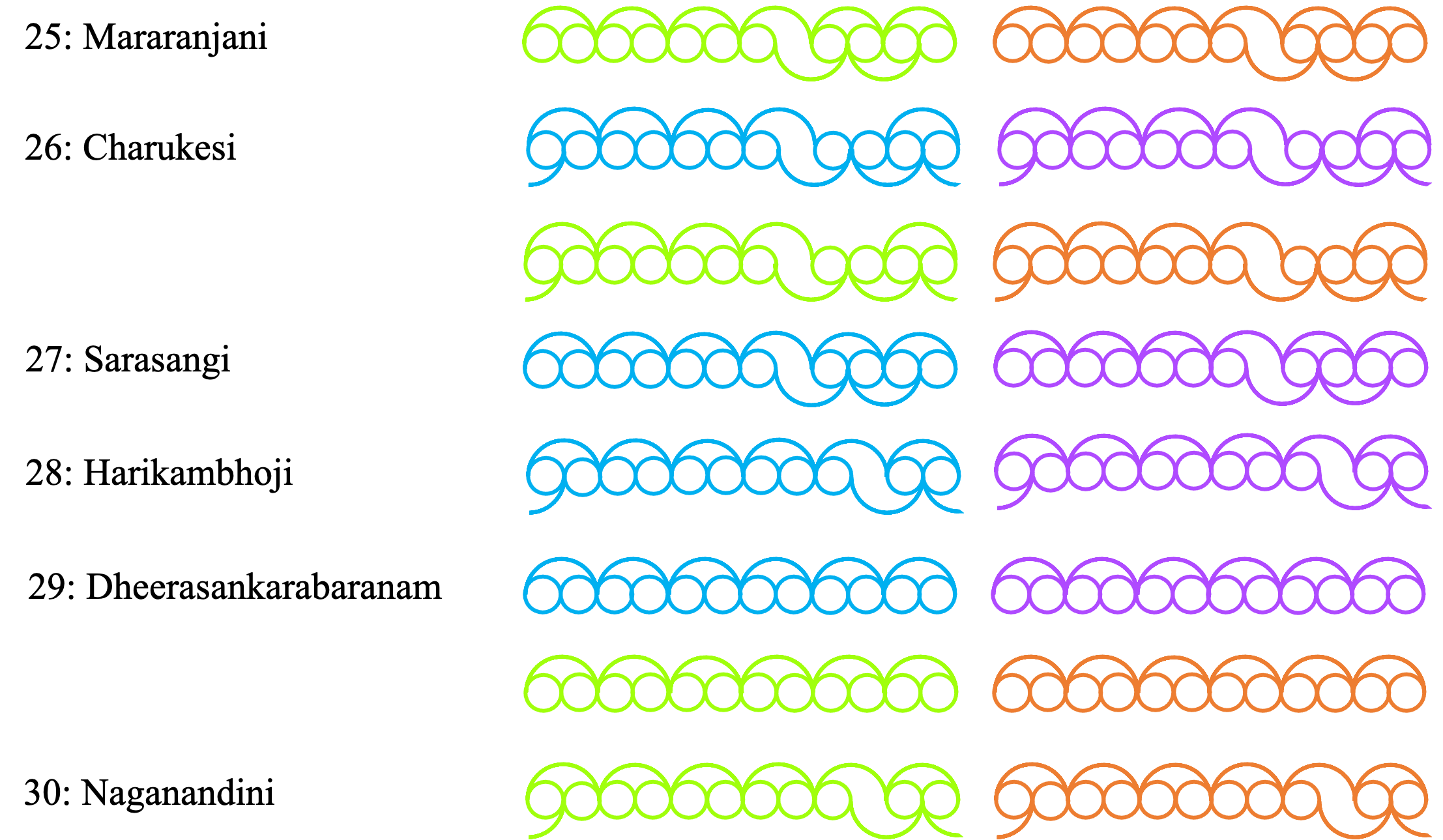}}}\hspace{5pt}
\caption{Representation of Mararanjani, Charukesi, Sarasangi, Harikambhoji, Dheerasankarabaranam, Naganandini by the circle-permutation diagram.} \label{Raga5}
\end{figure}

\begin{figure}[H]
\centering
{%
\resizebox*{14cm}{!}{\includegraphics{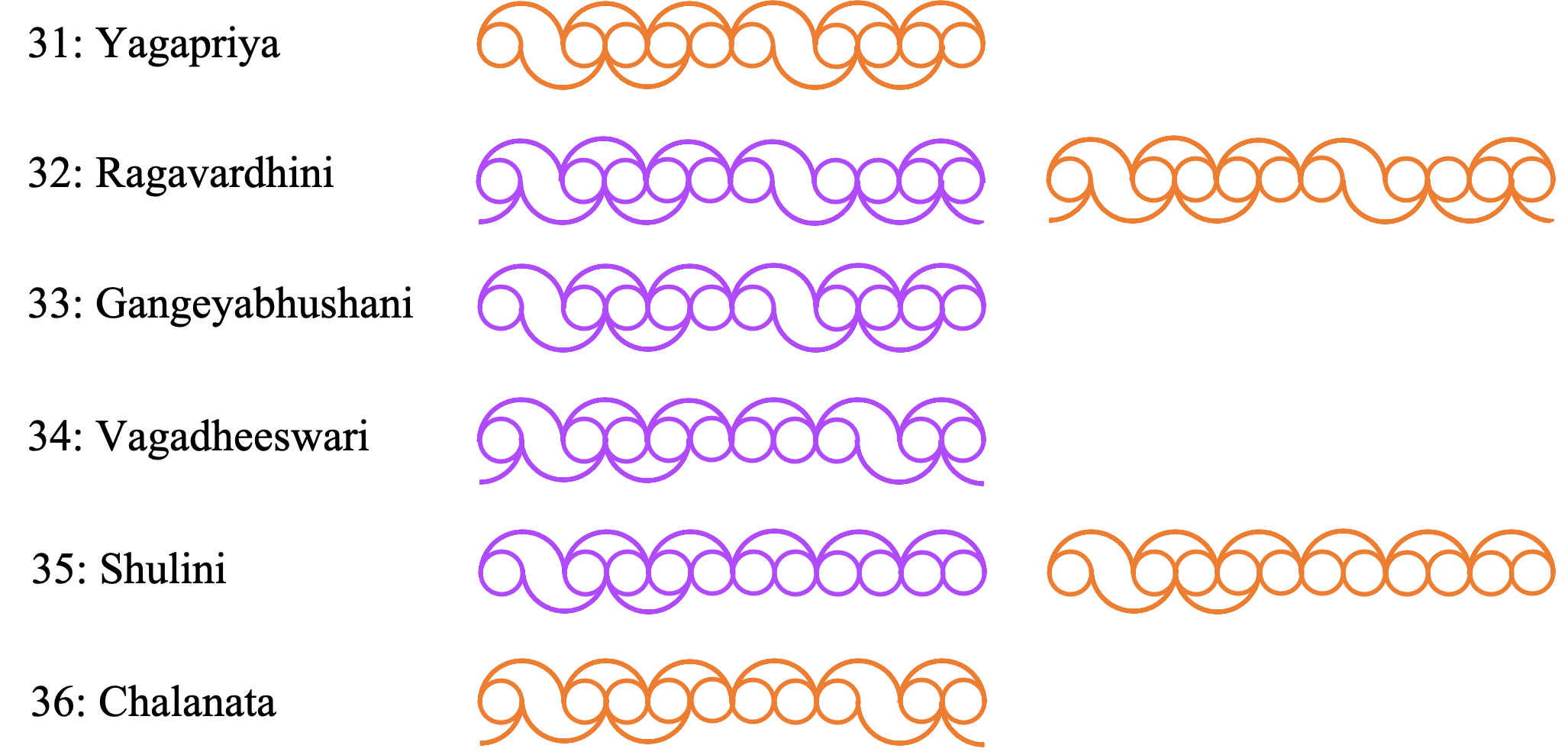}}}\hspace{5pt}
\caption{Representation of Yagapriya, Ragavardhini, Gangeyabhushani, Vagadheeswari, Shulini, Chalanata by the circle-permutation diagram.} \label{Raga6}
\end{figure}

\begin{figure}[H]
\centering
{%
\resizebox*{14cm}{!}{\includegraphics{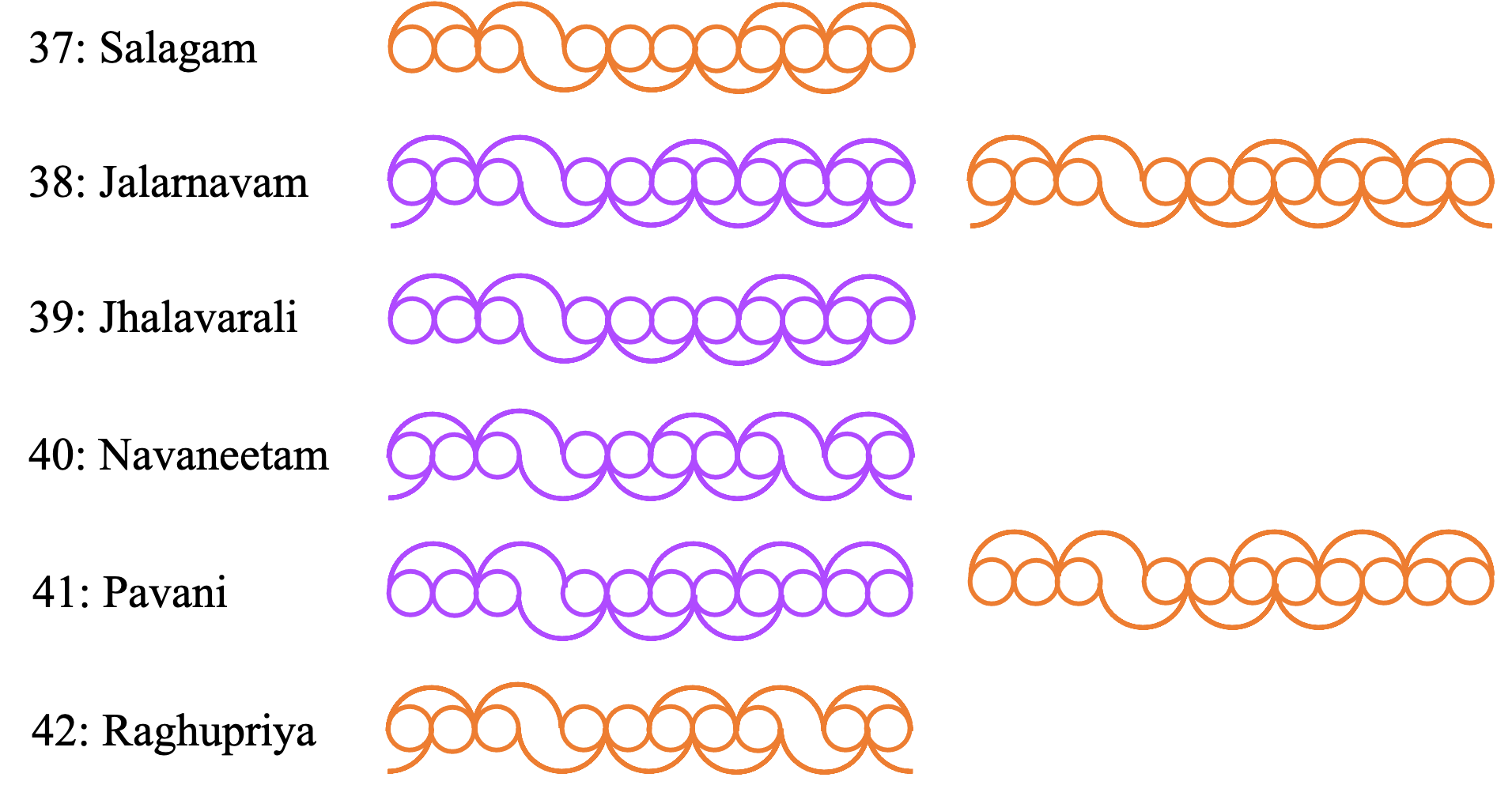}}}\hspace{5pt}
\caption{Representation of Salagam, Jalarnavam, Jhalavarali, Navaneetam, Pavani, Raghupriya by the circle-permutation diagram.} \label{Raga7}
\end{figure}

\begin{figure}[H]
\centering
{%
\resizebox*{14cm}{!}{\includegraphics{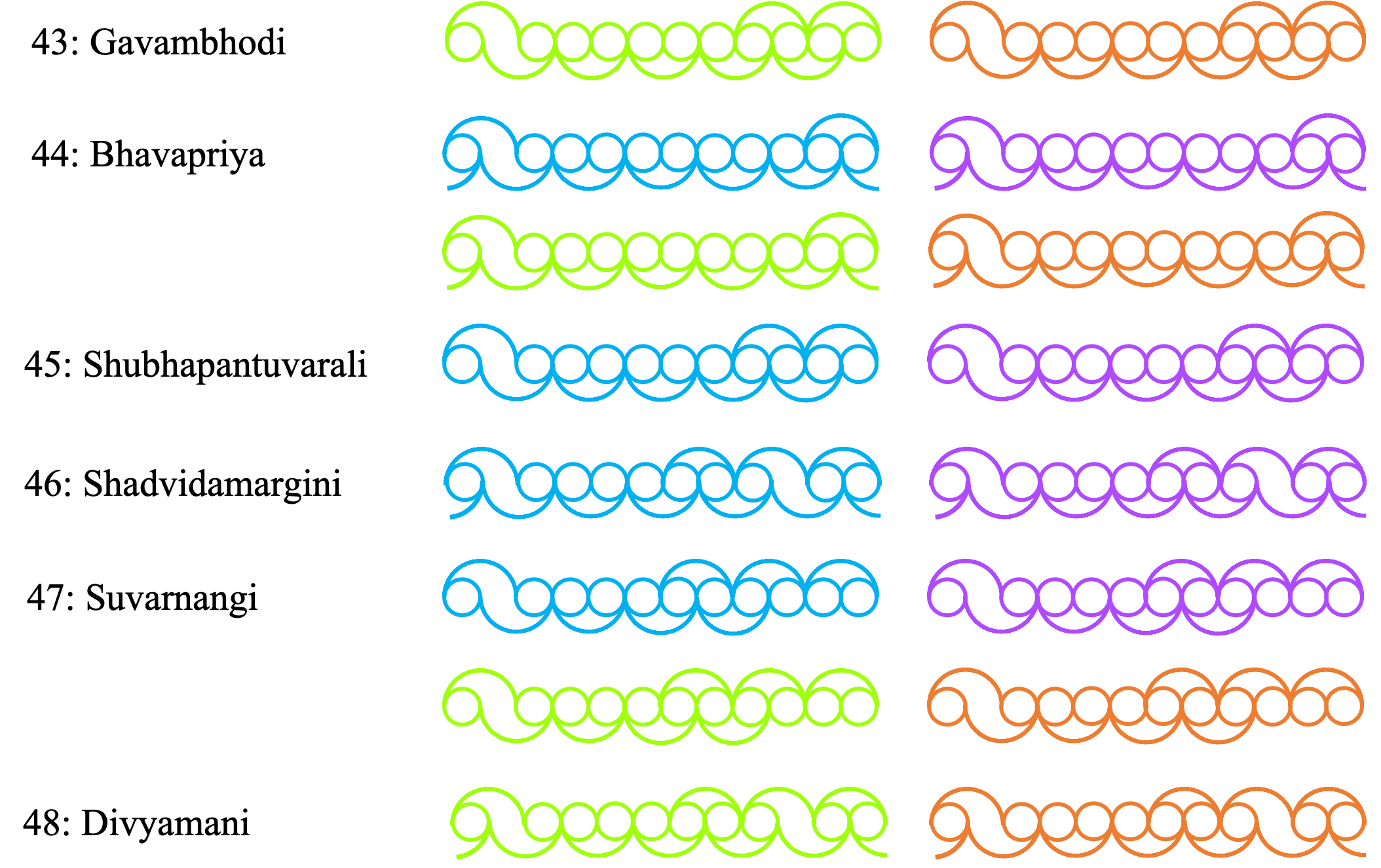}}}\hspace{5pt}
\caption{Representation of Gavambhodi, Bhavapriya, Shubhapantuvarali, Shadvidamargini, Suvarnangi, Divyamani by the circle-permutation diagram.} \label{Raga8}
\end{figure}

\begin{figure}[H]
\centering
{%
\resizebox*{14cm}{!}{\includegraphics{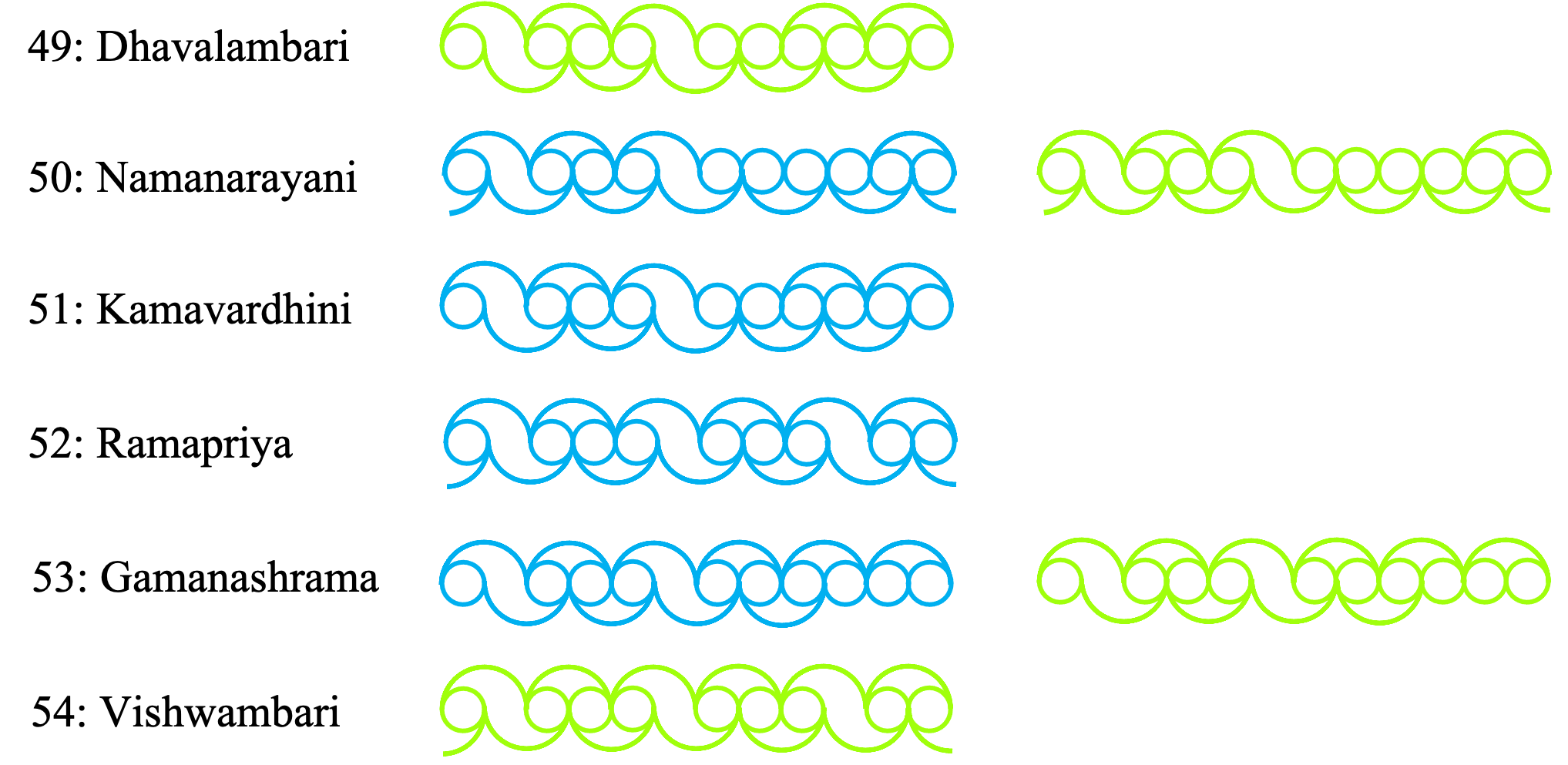}}}\hspace{5pt}
\caption{Representation of Dhavalambari, Namanarayani, Kamavardhini, Ramapriya, Gamanashrama, Vishwambari by the circle-permutation diagram.} \label{Raga9}
\end{figure}

\begin{figure}[H]
\centering
{%
\resizebox*{14cm}{!}{\includegraphics{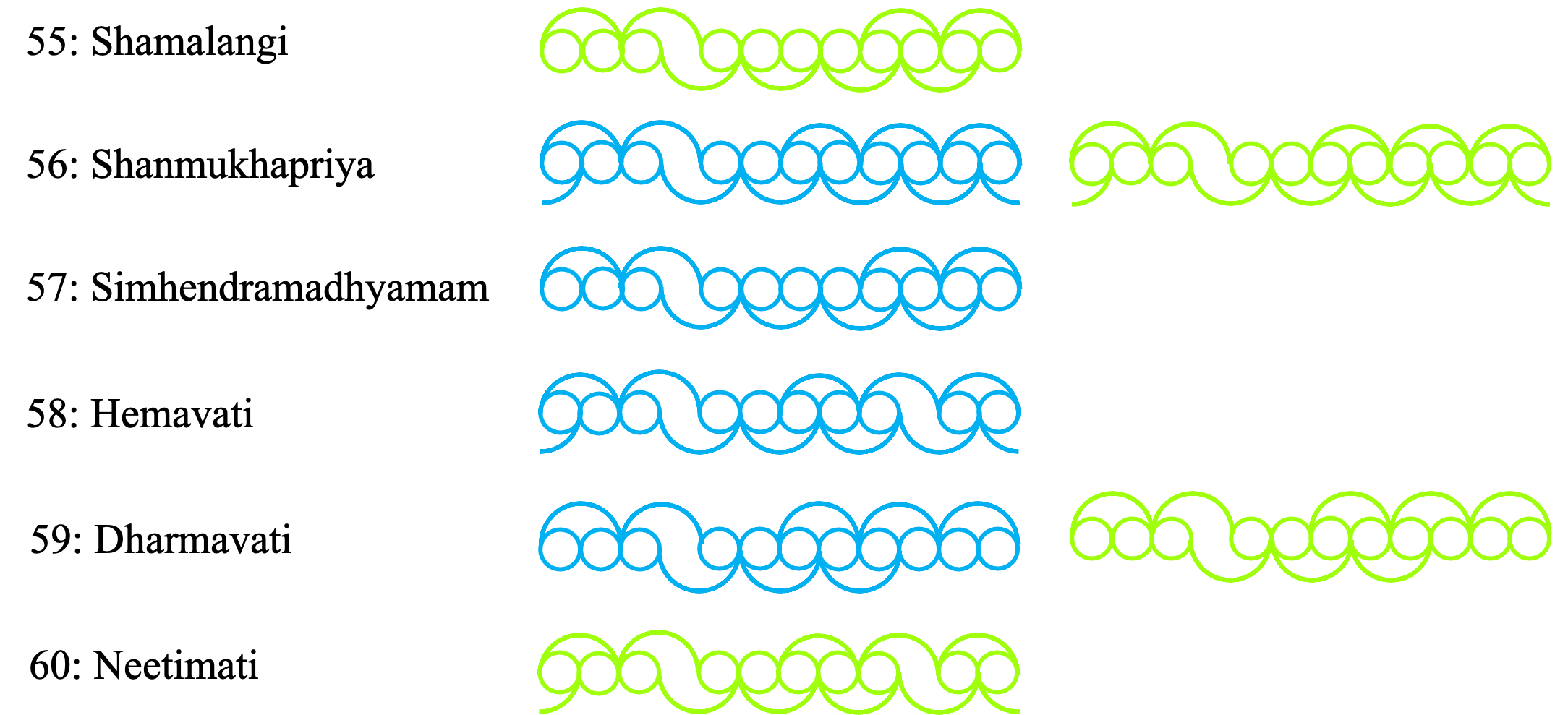}}}\hspace{5pt}
\caption{Representation of Shamalangi, Shanmukhapriya, Simhendramadhyamam, Hemavati, Dharmavati, Neetimati by the circle-permutation diagram.} \label{Raga10}
\end{figure}

\begin{figure}[H]
\centering
{%
\resizebox*{14cm}{!}{\includegraphics{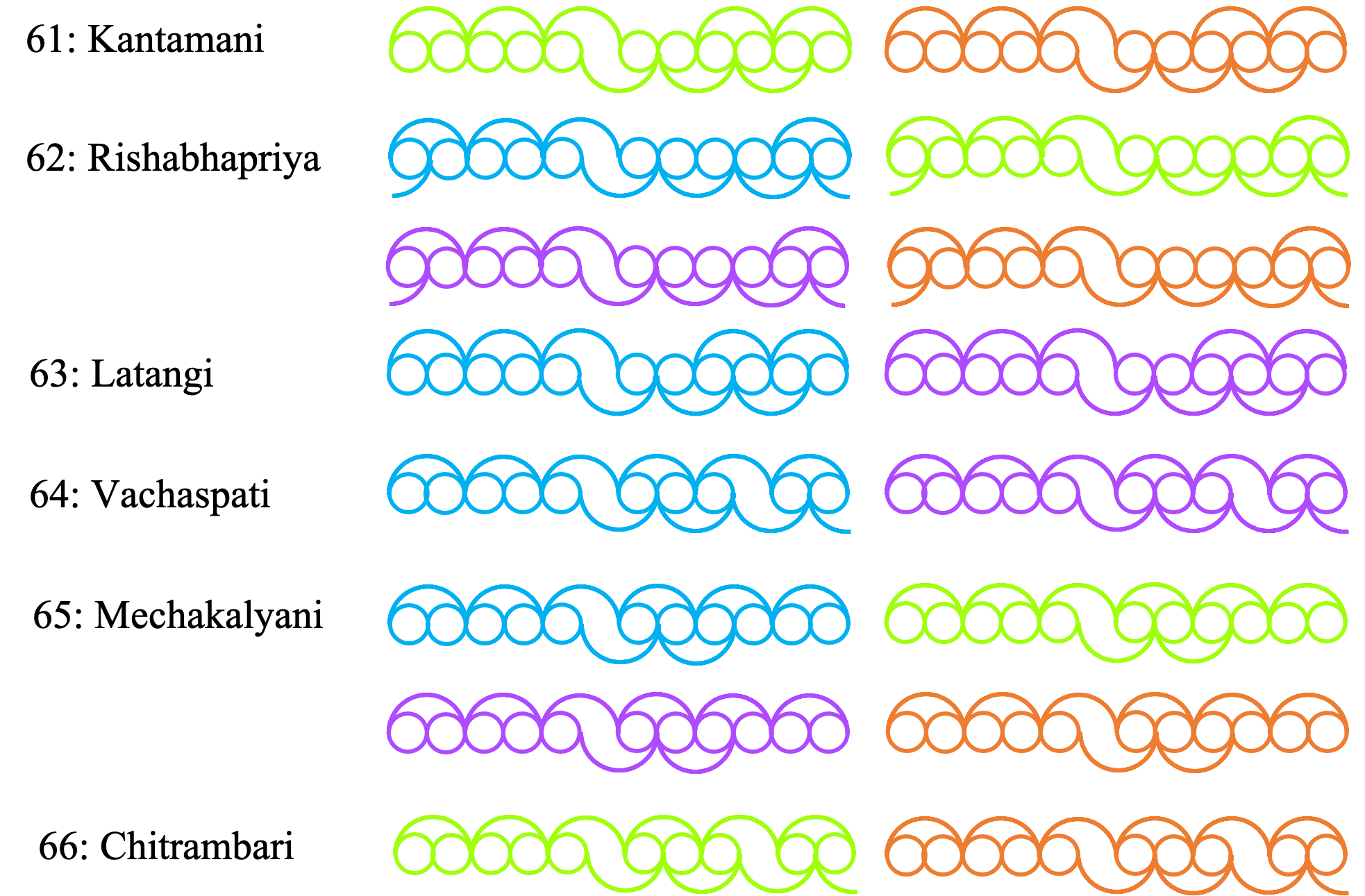}}}\hspace{5pt}
\caption{Representation of Kantamani, Rishabhapriya, Latangi, Vachaspati, Mechakalyani, Chitrambari by the circle-permutation diagram.} \label{Raga11}
\end{figure}

\newpage
\section{Conclusion}


We studied a connection between musical concepts in Western classical music and Melakarta raga that is a principal concept in Karnatak (south Indian) classical music, and we found the musical icosahedra that were proposed in our previous study can reveal the connection. Some kinds of musical icosahedra connecting various musical concepts in Western music were introduced in our previous study: chromatic/whole tone musical icosahedra (type 1, type 2, type 3, type 4), Pythagorean/whole tone musical icosahedra (type 1', type 2', type 3', type 4'), exceptional musical icosahedra (type ${\rm 1}^*$, type ${\rm 2}^*$, type ${\rm 3}^*$, type ${\rm 4}^*$). We introduced two kinds of permutations: inter-permutation and intra-permutation. The inter-permutaions connect the type 1 (type 2) and the type 3 (type 4), the type 1' (type 2') and the type 3' (type 4'), the type ${\rm 1}^*$ (type ${\rm 2}^*$) and the type ${\rm 3}^*$ (type ${\rm 4}^*$), and the intra-permutaions connect the type n and the type n'. We called the musical icosahedra characterizing these connections intermediate musical icosahedra. The intermediate musical icosahedra defined by the inter-permutations are characterized by 11 kinds of musical icosahedra, and the intermediate musical icosahedra defined by the intra-permutations are characterized by 13 kinds of musical icosahedra.

Next, we introduced a concept of the neighboring number that counts pairs of neighboring two tones in a given scale that neighbor each other on a given musical icosahedron. By combining linearly the neighboring numbers, we defined musical invariant. We found there exists a pair of a musical invariant and scales that is constant for the type 1 (type 2) and the type 3 (type 4) and the intermediate musical icosahedra from the type 1 (type 2) to the type 3 (type 4), the type 1' (type 2') and the type 3' (type 4') and for the intermediate musical icosahedra from the type 1' (type 2') to the type 3' (type 4'), and for the type ${\rm 1}^*$ (type ${\rm 2}^*$) and the type ${\rm 3}^*$ (type ${\rm 4}^*$) and the intermediate musical icosahedra from the type ${\rm 1}^*$ (type ${\rm 2}^*$) to the type ${\rm 3}^*$ (type ${\rm 4}^*$). We visualized them by coloring a diagram made by 12 circles, 11 half-circles, and 2 quarter-circles. We also found there exists a pair of a musical invariant and scales that is constant for the type n, the type n', and the intermediate musical icosahedra from the type n to the type n'. We visualized them by using intersections of a diagram made by 6 ellipses and 12 circles.

Last, we introduced a concept of the permutation-extension that is an extension of a given scale by the inter-permutations of a given musical icosahedron. Then, we showed the permutation-extension of the $C$ major scale by Melakarta raga musical icosahedra that are four of the intermediate musical icosahedra from the type 1 chromatic/whole tone musical icosahedron to the type 1' Pythagorean/whole tone musical icosahedron, is a set of all the scales included in Melakarta raga. We found there exists a musical invariant that is constant for all the musical icosahedra corresponding to the scales of Melakarta raga, and we obtained a diagram representation of those scales characterizing the musical invariant and having 15 circles totally.

Imai, Y. Dellby, S. C., and Tanaka, N., ``General Theory of Music by Icosahedron 1: A bridge between "artificial" scales and "natural" scales, Duality between chromatic scale and Pythagorean chain, and Golden Major Minor Self-Duality", arXiv:2103.10272.

Imai, Y., ``General Theory of Music by Icosahedron 2: Analysis of musical pieces by the exceptional musical icosahedra", arXiv:2108.10294.

Powers, H. S. and Widdess, R. ``R${\rm \bar{a}}$ga" from ``India, subcontinent of" by Qureshi, R. \emph{et al}., Grove Music, 2020.

Krishna, T. M., ``KARNATIK MUSIC: SVARA, GAMAKA, PHRASEOLOGY AND RAGA IDENTITY", Proc. of the 2nd CompMusic Workshop, 2012.

Mathur, A. \emph{et al}., ``Emotional responses to Hindustani raga music: the role of musical structure", Frontiers in Psychology, Emotion Science, 2015.

Gitanjali, B., ``EFFECT OF THE KARNATIC MUSIC RAGA ``NEELAMBARI" ON SLEEP ARCHITECTURE", Indian J Physiol Pharmacal, 42(1), 1998.

Sairam, T.V., ``MELODY AND RHYTHM– 'Indianness' in Indian music and music therapy", Music Therapy Today, 7(4), 2004.

Pandey, G., Mishra, C, and Ipe, P., ``TANSEN: A System for Automatic Raga Identification", Proceedings of the 1st Indian International Conference on Artificial Intelligence, IICAI, 2003.

Shetty, S. and Achary,  K.K., ``Raga Mining of Indian Music by Extracting Arohana-Avarohana Pattern", International Journal of Recent Trends in Engineering, 1(1), 2009.

Sridhar, R. and Geetha, T. V., ``Raga Identification of Carnatic music for Music
Information Retrieval", International Journal of Recent Trends in Engineering, 1(1), 2009.

Ross, J. C. and Rao, P., ``DETECTION OF \emph{RAGA}-CHARACTERISTIC PHRASES FROM HINDUSTANI CLASSICAL MUSIC AUDIO", Proc. of the 2nd CompMusic Workshop, 2012.

Rao, P. \emph{et al}., ``Classification of Melodic Motifs in Raga Music with Time-series Matching", Journal of New Music Research, 43(1), 2014.

\end{document}